\documentclass[10pt,letterpaper]{article}
\usepackage{subfig}
\usepackage{opex3}
\usepackage{float}
\usepackage[utf8]{inputenc}
\usepackage[nice]{nicefrac}
\usepackage{amssymb} 
\usepackage{amsmath} 
\usepackage{amsfonts}
\usepackage{gensymb}
\usepackage{array}
\usepackage{multirow}

\begin{document}

\title{First on-sky results of the CO-SLIDAR $C_n^2$ profiler}

\author{Juliette Voyez,$^{1}$ Clélia Robert,$^{1,*}$ Jean-Marc Conan,$^{1}$ Laurent Mugnier,$^{1}$ Etienne Samain,$^{2}$ Aziz Ziad,$^{3}$}

\address{$^{1}$ONERA, The French Aerospace Lab, Châtillon, France \\ $^{2}$ Laboratoire Géoazur, Université de Nice-Sophia Antipolis, CNRS, OCA, Caussols, France\\ $^{3}$Laboratoire
  Lagrange, Université de Nice-Sophia Antipolis, CNRS, OCA, Nice, France}

\email{$^*$clelia.robert@onera.fr} 



\begin{abstract}
  COupled SLope and scIntillation Detection And Ranging (CO-SLIDAR) is a recent
  profiling method of the vertical distribution of atmospheric turbulence
  strength ($C_n^2$ profile). It takes advantage of correlations of slopes and of
  scintillation, both measured with a Shack-Hartmann wavefront sensor on a binary star.
  In this paper, we present the improved CO-SLIDAR reconstruction method of
  the $C_n^2$ profile and the first on-sky results of the CO-SLIDAR profiler. We examine CO-SLIDAR latest
  performance in simulation, taking into account the detection noise bias and
  estimating error bars along with the turbulence profile. The estimated
  $C_n^2$ profiles demonstrate the accuracy of the CO-SLIDAR method,
  showing sensitivity to both low and high altitude turbulent layers.
  CO-SLIDAR is tested on-sky for the first time, on the $1.5$~m MeO
  (Métrologie Optique) telescope at Observatoire de la Côte d'Azur (France).
  The reconstructed profiles are compared to turbulence profiles estimated from
  meteorological data and a good agreement is found. We discuss CO-SLIDAR's
  contribution in the $C_n^2$ profilers' landscape and we propose some improvements of the
  instrument. 
\end{abstract}


\section{Introduction}

The vertical distribution of atmospheric turbulence strength, known as the $C_n^2$
profile, is a key-point in the development of next-generation Adaptive Optics
(AO) systems. Wide Field AO (WFAO) such as Laser Tomographic AO (LTAO),
Multi-Conjugated AO (MCAO) and Multi-Object AO (MOAO) are based on a
tomographic reconstruction of the atmospheric turbulence. They require a
detailed knowledge of the $C_n^2$ profile for AO design and performance
evaluation purposes~\cite{lelouarn-a-00, fusco-a-01,costille-p-11} and for the
forecast of turbulence conditions~\cite{masciadri-a-13} to perform flexible
scheduling of the observations with the instruments. $C_n^2$ profiles with a
sub-kilometric vertical resolution in the $\left[0 \;; \; 20\right]$~km
vertical range are necessary for these objectives~\cite{costille-p-11}. 
\newpage
Many $C_n^2$ profilers have been developed through the last decades. SLODAR
(SLOpe Detection And Ranging)~\cite{wilson-a-02, butterley-a-06} uses
correlations of slopes measured on a binary star with a Shack-Hartmann wavefront sensor (SH) to
estimate the turbulent profile. Correlations of slopes allow a high sensitivity
to ground and low altitude layers, but the performance is degraded at high
altitudes. Other methods use correlations of scintillation. Among them
  SCIDAR (SCIntillation Detection And Ranging)~\cite{rocca-a-74}, which works on a double star, and MASS (Multiple Aperture Scintillation
Sensor)~\cite{tokovinin-a-03b}, which uses a single star but only measures six
turbulent layers. These techniques work better at high altitude and do not
allow the estimation of near-ground turbulence, because they
employ scintillation, whose variance is proportional to the altitude raised
to the power $5/6^{\mathrm{th}}$. The Generalized SCIDAR
(G-SCIDAR)~\cite{avila-a-97, fuchs-a-98} is an improvement of the SCIDAR that allows
the method to be sensitive to the ground layer. The plane of the detector
is the optical conjugate of a plane at a distance $h_{\mathrm{gs}}$ which is
below the telescope pupil, and this extra-propagation length enables
G-SCIDAR to detect near-ground turbulence.

CO-SLIDAR~\cite{vedrenne-a-07} is a recent $C_n^2$ profiler, combining
sensitivity to both low and high altitude layers, which jointly uses correlations
of slopes and of scintillation, both measured on a binary star with a
SH. CO-SLIDAR's first validation in simulation have been presented
in~\cite{vedrenne-a-07}. The next step is a full experimental validation
of the concept. To prepare the experiment, some experimental tests were
performed in~\cite{voyez-p-11} and an end-to-end simulation of the
estimation of the $C_n^2$ profile in a practical case was detailed in~\cite{voyez-p-12}.

This paper presents two improvements on the CO-SLIDAR reconstruction method and the first
on-sky results of the CO-SLIDAR $C_n^2$
profiler. Detection noises are now taken into account and the resulting bias
is subtracted in the reconstruction process. Error bars on the profile are now
computed along with the estimated profile. The latest performance of CO-SLIDAR
is first described in a simulated case. CO-SLIDAR is then tested on-sky, on
the $1.5$~m MeO telescope.
Images on a binary star are acquired to extract slope and
scintillation data. Their correlations are computed so as to estimate $C_n^2$
profiles. The CO-SLIDAR profiles are estimated with
their error bars. Results are compared to those obtained from correlations of slopes only
or of scintillation only. Then, the CO-SLIDAR profiles are compared to profiles estimated from
meteorological data, from NCEP (National Centers for Environmental
Prediction)/NCAR (National Centers for Atmospheric Research) Reanalysis data
base~\cite{hach-a-12}. We finally discuss CO-SLIDAR's contribution as a new
$C_n^2$ profiler and we suggest some improvements.        

This paper is organized as follows. In Section~\ref{sec:coslidar}, we recall
CO-SLIDAR's principle and formalism and we describe the improvements on the reconstruction. In
Section~\ref{sec:simu} we show the latest performance of the
CO-SLIDAR in a simulated case. Section~\ref{sec:on_sky_results} is dedicated to the on-sky experiment and results. In
Section~\ref{sec:position} we discuss CO-SLIDAR's contribution in the $C_n^2$
profilers' landscape. Our conclusions and perspectives are given in Section~\ref{sec:conclusion}.

\section{CO-SLIDAR's principle and formalism}\label{sec:coslidar}
CO-SLIDAR is based on correlations of slopes and correlations of scintillation
measured with a SH on a binary star. In Subsection~\ref{sec:principle}, we first recall
CO-SLIDAR's principle and theory. We present the direct problem in
Subsection~\ref{sec:direct_problem}. In Subsection~\ref{sec:ml_solution}, we describe the
Maximum Likelihood inversion together with the debiasing process and we
detail the computation of the error bars. In Subsection~\ref{sec:map_solution}
we present the Maximum \textit{A Posteriori} inversion. Finally, in
Subsection~\ref{sec:alt_res}, we recall commonly used
altitude resolution formula and explain which one we adopt for CO-SLIDAR.  

\subsection{CO-SLIDAR's principle}\label{sec:principle}
Given a double star with angular separation $\mathrm{\pmb{\theta}}$ in
the field of view, the SH data at a given time $t$ are a set of wavefront
slopes and intensity fluctuations (scintillation) per star. 

For a star at angular position $\mathrm{\pmb{\alpha}}$, the slope
is measured in subaperture $\left(u,v \right)$, where
$u$ and $v$ denote the position of the subaperture in the SH array in the
horizontal ($x$) and vertical ($y$) directions. The slope,
denoted $\mathbf{s}_{u,v}\left(\mathrm{\pmb{\alpha}}\right)$, is a bidimensional
vector with components $s_{u,v}^k$ along the $k$-axis, $k \in \{x,y\}$. The star
intensity in the subaperture $\left(u,v\right)$, $i_{u,v}\left(\mathrm{\pmb{\alpha}}\right)$, leads
to the relative intensity fluctuation $\delta
i_{u,v}\left(\mathrm{\pmb{\alpha}}\right)=\frac{i_{u,v}\left(\mathrm{\pmb{\alpha}}\right)-\langle
  i_{u,v}\left(\mathrm{\pmb{\alpha}}\right)\rangle}{\langle
  i_{u,v}\left(\mathrm{\pmb{\alpha}}\right)\rangle}$ where $\langle
i_{u,v}\left(\mathrm{\pmb{\alpha}}\right)\rangle$ is the temporal average of
$i_{u,v}\left(\mathrm{\pmb{\alpha}}\right)$. 

Spatial correlations of slopes $\langle
s^k_{u,v}\left(\mathrm{\pmb{\alpha}}\right)s^l_{u+\delta u, v+\delta
  v}\left(\mathrm{\pmb{\alpha}} + \mathrm{\pmb{\theta}}\right)
\rangle$ and spatial correlations of scintillation $\langle \delta i_{u,v}\left(\mathrm{\pmb{\alpha}}\right) \delta
i_{u+\delta u, v+\delta v}\left(\mathrm{\pmb{\alpha}} + \mathrm{\pmb{\theta}}\right) \rangle$,
calculated between subapertures $\left(u,v\right)$ and $\left(u+\delta u, v+\delta v\right)$, of
separation vector $\pmb{\rho}=\left(\delta u, \delta v\right)$, are directly related to integrals
of the $C_n^2\left(h\right)$ weighted by theoretical functions $W_{ss}^{kl}$ and
$W_{ii}$ and are written as:
\begin{eqnarray}
  \label{eq:corr_slo}
  \langle s^k_{u,v}\left(\mathrm{\pmb{\alpha}}\right) s^l_{u+\delta u,
    v+\delta v}\left(\mathrm{\pmb{\alpha}} + \mathrm{\pmb{\theta}}\right)
  \rangle &=&
  \int_{0}^{+\infty}C_{n}^{2}\left(h\right)W_{ss}^{kl}\left(h,\mathrm{\pmb{\rho}},
    \mathrm{\pmb{\theta}}\right)\,\mathrm{d}h,\\
  \label{eq:corr_sci}
  \langle \delta i_{u,v}\left(\mathrm{\pmb{\alpha}}\right) \delta i_{u+\delta
    u, v+\delta v}\left(\mathrm{\pmb{\alpha}} + \mathrm{\pmb{\theta}}\right)
  \rangle &=&
  \int_{0}^{+\infty}C_{n}^{2}\left(h\right)W_{ii}\left(h,\mathrm{\pmb{\rho}}, \mathrm{\pmb{\theta}}\right)\,\mathrm{d}h,
\end{eqnarray}
where $\langle \rangle$ denotes the averaging over time series. These
expressions are derived from the theory of anisoplanatism effect in the 
weak perturbation regime~\cite{robert-a-06}. $W_{ss}^{kl}$ and
$W_{ii}$ are weighting functions that depend on the turbulence spectrum (which
is commonly described using a von K\'arm\'an spectrum), the altitude $h$, the separation vector $\mathrm{\pmb{\rho}}$ between the
subapertures, and the star separation $\mathrm{\pmb{\theta}}$.

In CO-SLIDAR, we compute both cross-correlations, combining the measurements
on the two stars, and auto-correlations, corresponding to the measurements on
a single star (case $\mathrm{\pmb{\theta}}=0$ in Eqs.~(\ref{eq:corr_slo})
and~(\ref{eq:corr_sci})). Correlations of slopes bring sensitivity to ground and
low altitude layers, whereas correlations of scintillation mainly give
sensitivity to high altitude layers.

\subsection{Direct problem}\label{sec:direct_problem}
In CO-SLIDAR, we currently only exploit correlations of $x$-slopes, of
$y$-slopes and of scintillation for reason of computing time and because
cross-correlations between $x$ and $y$-slopes are generally weaker.
Correlations are averaged over all pairs of subapertures with given separation
and represented as auto- and cross-correlation maps, with horizontal and vertical dimensions $2
\times n-1$, where $n$ is the number of subapertures across the telescope diameter. Then, one pixel of these maps represents
the pseudo-measurement that can be written, respectively for correlations of
slopes and of scintillation, as:
\begin{eqnarray}
  C_{ss}^{kk}\left(\delta u, \delta v, \mathrm{\pmb{\theta}} \right) &=& \frac{\sum_{u,v}\langle s_{u,v}^{k} s_{u+\delta u, v+
      \delta v}^{k} \rangle
    \left(\mathrm{\pmb{\theta}}\right)}{N\left(\delta u, \delta
      v\right)}, \quad k \in \{x,y\},\label{eq:corr_slo2}\\
  C_{ii}\left(\delta u, \delta v,\mathrm{\pmb{\theta}} \right) &=& \frac{\sum_{u,v}\langle \delta
    i_{u,v} \delta i_{u+\delta u, v+\delta v} \rangle
    \left(\mathrm{\pmb{\theta}}\right)}{N\left(\delta u, \delta
      v\right)}. \label{eq:corr_sci2}
\end{eqnarray} 
$\sum_{u,v}$ denotes a summation over all pairs of subapertures with
separation $\mathrm{\pmb{\rho}}=\left(\delta u, \delta v\right)$ and
$N\left(\delta u, \delta v\right)$ represents the number of pairs of such subapertures. The
pseudo-measurements given by Eqs.~(\ref{eq:corr_slo2}) and~(\ref{eq:corr_sci2})
are then stacked into a single vector $\mathbf{C_{mes}}$, related to the discretized $C_n^2$ profile
at different altitudes $\mathbf{C_n^2}$, by the following linear relationship:
\begin{equation}
  \label{eq:direct_problem}
  \mathbf{C_{mes}} = M \mathbf{C_n^2} + \mathbf{C_d} + \mathbf{u}.
\end{equation}
$M$ is the matrix of the weighting functions $W_{ss}^{kk}$ and $W_{ii}$ of
Eqs.~(\ref{eq:corr_slo}) and~(\ref{eq:corr_sci}).
Because slope and scintillation data are affected by detection noises, the
pseudo-measurements $\mathbf{C_{mes}}$ are biased with the averaged
correlations of these noises $\mathbf{C_d}$.
As we estimate the correlations from a finite number of frames, $\mathbf{u}$
represents a convergence noise, which we assume to be Gaussian in the following.

\subsection{Maximum Likelihood solution}\label{sec:ml_solution}

The $C_n^2$ profile is retrieved by minimizing the following Maximum
Likelihood (ML) criterion, which is the opposite of the data log-likelihood~\cite{mugnier-b-08}:
\begin{equation}
  \label{eq:ML_J} 
  J_{\mathrm{ML}}\left(\mathbf{C_n^2}\right) = \left(\mathbf{{C}_{mes}} - \mathbf{C_d} - M \mathbf{C_n^2}\right)^{\mathrm{T}} C_{\mathrm{conv}}^{-1}
  \left(\mathbf{C_{mes}} - \mathbf{C_d} - M \mathbf{C_n^2}\right).
\end{equation}
Under the adopted Gaussian assumption for the convergence noise, this
criterion is quadratic and its unconstrained minimization thus has an analytical solution. Yet, as the
$C_n^2$ is always non-negative, we minimize $J_{\mathrm{ML}}$ under positivity
constraint, using a VMLM~-~B (Variable Metric with Limited
  Memory-and Bounds) algorithm~\cite{thiebaut-p-02}. 
The covariance matrix of the convergence noise $\mathbf{u}$,
$C_{\mathrm{conv}}=\langle \mathbf{u} \mathbf{u}^{\mathrm{T}} \rangle$, is
deduced from an analytical expression depending on the theoretical correlations,
which are in practice approximated with the pseudo-measurements~\cite{vedrenne-t-08}.
The idea of the derivation of $C_{\mathrm{conv}}$ can be gained from examining
the scalar case of a single variable. To this aim, Appendix~\ref{sec:appendix}
gives the calculation of the variance of the convergence noise for a centered Gaussian
random variable.

Two methods are considered to take into account the bias due to detection
noises $\mathbf{C_d}$.
Assuming that the noises are not correlated between the two directions of
observation and between different subapertures, only the variances of slopes
and of scintillation are biased. These variances are averaged over all
subapertures and represent the central point of the auto-correlation maps. Three new parameters,
\textit{i.e.} the variances of the noises on $x$-slopes, $y$-slopes and
scintillation, can be estimated jointly with the $C_n^2$ profile,
without changing the ML criterion given by Eq.~(\ref{eq:ML_J}). 
Another option is to exclude the variances of slopes and of scintillation from the vector
$\mathbf{C_{mes}}$. These two methods will be tested and compared further.

Using Eq.~(\ref{eq:ML_J}), the error covariance matrix of the ML estimation can be
shown~\cite{mugnier-b-08} to be $\mathrm{cov}\left(\mathbf{\hat{C}_{n_{\mathrm{ML}}}^2}\right) = \left(M^{\mathrm{T}}C_{\mathrm{conv}}^{-1}
  M\right)^{-1}$ and can be used to obtain error bars on the restored profile. Indeed, the
diagonal values of this matrix represent the variances of the estimates of the
$C_n^2$ as a function of altitude, from which $3\sigma$ error bars are easily
computed and will be considered as an upper
bound of the error on the reconstructed profile. 

\subsection{Maximum \textit{A Posteriori} solution}\label{sec:map_solution}
As the $C_n^2$ profile is supposed to be smooth near the ground, we can also minimize a Maximum \textit{A Posteriori} (MAP)
metric~\cite{mugnier-b-08} composed of the ML criterion $J_{\mathrm{ML}}$
and a regularization metric, denoted hereafter by $J_{\mathrm{p}}$ and
designed to enforce smoothness of the $C_n^2$ profile:
\begin{equation}
  \label{eq:MAP_J} 
  J_{\mathrm{MAP}}\left(\mathbf{C_n^2}\right) = \left(\mathbf{{C}_{mes}} - \mathbf{C_d} - M \mathbf{C_n^2}\right)^{\mathrm{T}} C_{\mathrm{conv}}^{-1}
  \left(\mathbf{C_{mes}} - \mathbf{C_d} - M \mathbf{C_n^2}\right) + J_{\mathrm{p}}\left(\mathbf{C_n^2}\right).
\end{equation}
In this paper, we take:
\begin{equation}
  \label{eq:Jp}
  J_{\mathrm{p}}\left(\mathbf{C_n^2}\right)= \beta ||\nabla \mathbf{C_n^2}||^2,
\end{equation}
where $\beta$ is a regularization parameter and $\nabla$ represents the
gradient operator. With this kind of regularization, the constraint applies
mainly on strong turbulent layers, located near the ground.

Later in this article, we will use either the ML or the MAP solution.

\subsection{Altitude resolution}\label{sec:alt_res}
In cross-correlation methods, simple geometrical rules are
used to define the altitude resolution $\delta h$ of the instrument. For
SLODAR, we have~\cite{wilson-a-02}:
\begin{equation}
  \label{eq:res_slodar}
  \delta h\simeq \frac{d_{\mathrm{sub}}}{\theta},
\end{equation}
where $d_{\mathrm{sub}}$ is the subaperture diameter. For G-SCIDAR, we have~\cite{avila-a-97, prieur-a-01}:
\begin{equation}
  \label{eq:res_gscidar}
  \delta h \simeq 0.78 \frac{\left(\lambda |h-h_{\mathrm{gs}}|\right)^{\nicefrac{1}{2}}}{\theta},
\end{equation}
where $\lambda$ is the observation wavelength and $h_{\mathrm{}gs}$ the extra-propagation length. 
For both methods, the maximum altitude of sensitivity $H_{\mathrm{max}}$ is
constrained by the telescope diameter $D$ and the star separation $\theta$ and is
written as~\cite{wilson-a-02}:
\begin{equation}
  \label{eq:alt_max}
  H_{\mathrm{max}}\simeq \frac{D}{\theta}.
\end{equation}

In this paper, the altitude sampling and the maximum altitude of sensitivity of the
CO-SLIDAR will be given by Eqs.~(\ref{eq:res_slodar})
and~(\ref{eq:alt_max}) as we use a SH, like in the SLODAR method.

\section{Simulation results}\label{sec:simu}
In this section, we present the results of a numerical simulation in a practical
astronomical case. In Subsection~\ref{sec:data_simu} we describe the
simulation and we show the resulting correlation maps. In
Subsection~\ref{sec:profiles_simu}, we estimate the $C_n^2$ profiles and discuss the results.

\subsection{Simulated data and correlation maps}\label{sec:data_simu}
We consider a $30\times 30$ SH, conjugated with a telescope
of diameter $D=1.5$~m with a central obscuration of $30$~\%. The subaperture
diameter is $d_{\mathrm{sub}}=5$~cm. The object is a binary star with
separation $\theta=20$'', observed at $\lambda=0.55$ $\mathrm{\mu}$m. The two stars have a difference of one magnitude and
the fluxes are about $120$ and $300$ photons per subaperture and per frame. We
use a $C_n^2$ profile sampled on $32$ layers every $625$~m. This profile corresponds to a Fried parameter
$r_{0_{\mathrm{th}}} \simeq 5.5$~cm, and a variance of log-amplitude $\sigma_{\chi_{\mathrm{th}}}^2 \simeq
0.023$.

\begin{figure}[htbp]
  \begin{center}\leavevmode
    \subfloat[]{
      \includegraphics[width=0.4\linewidth]{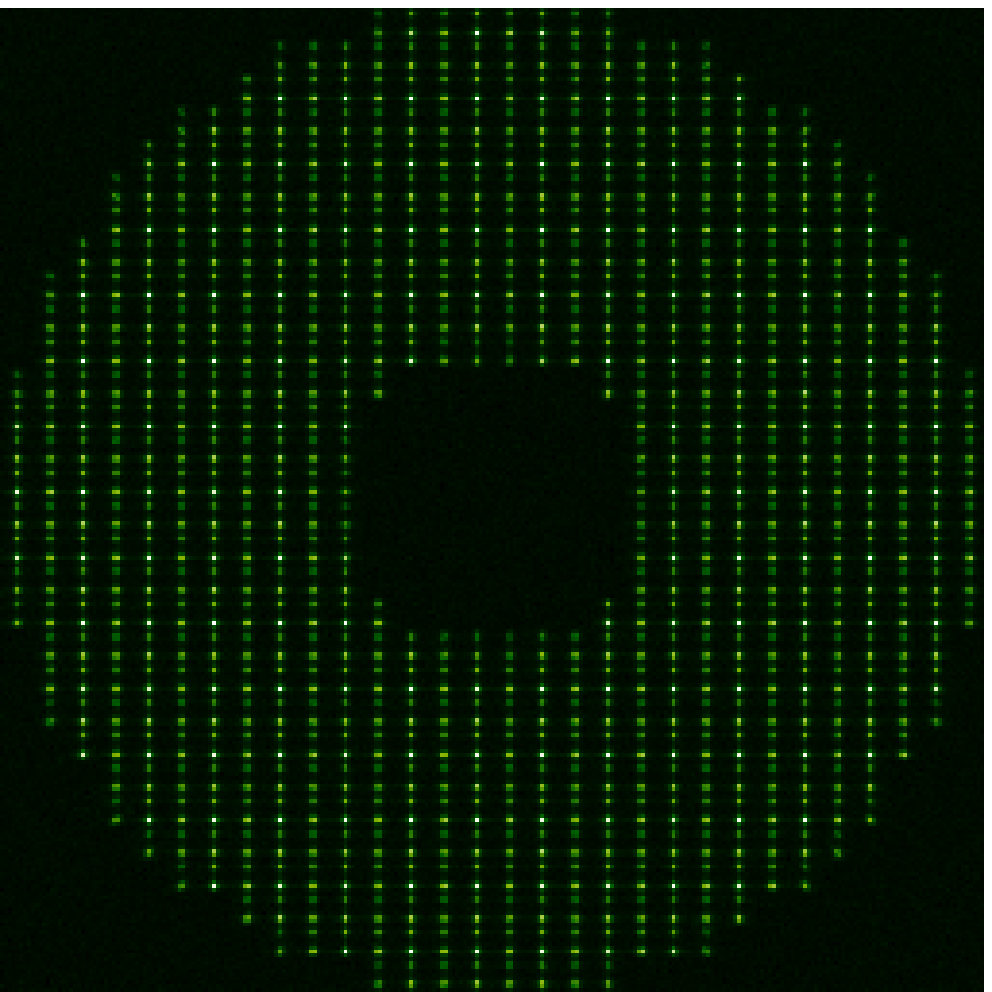}}
    \subfloat[]{
      \includegraphics[width=0.4\linewidth]{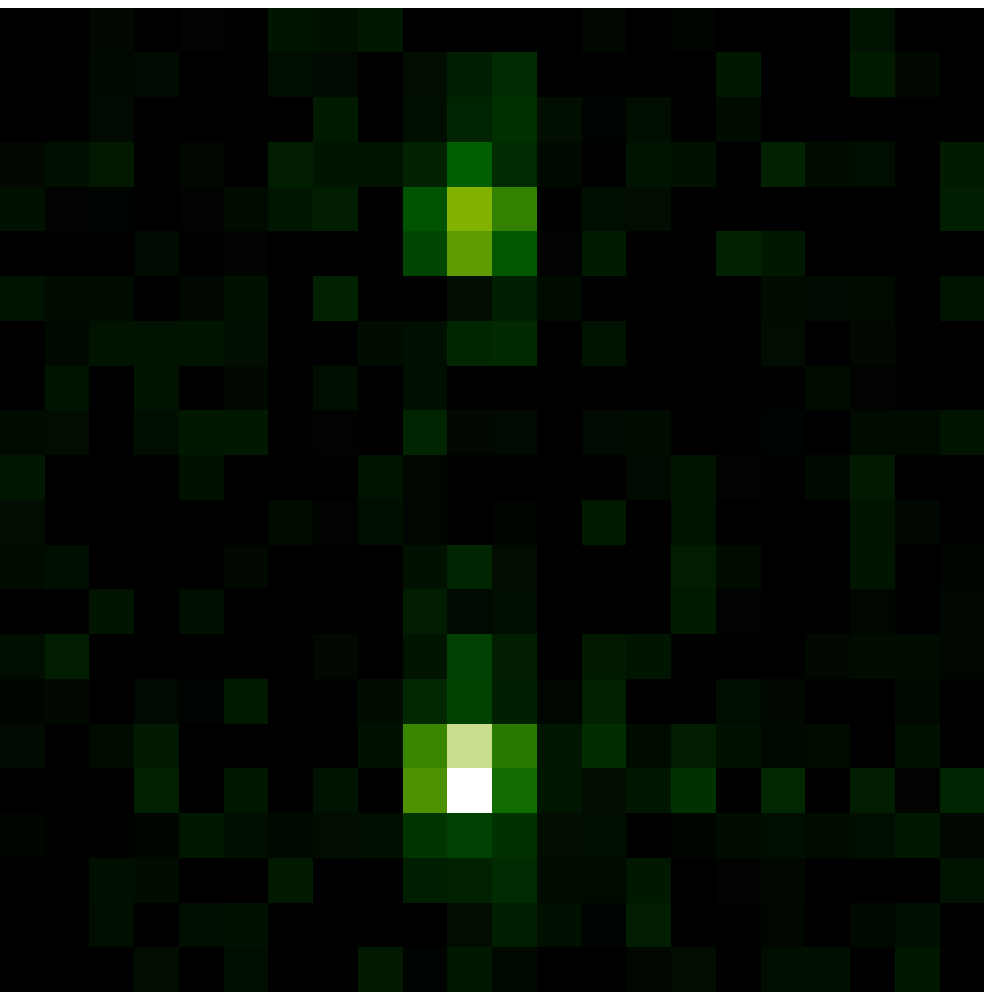}}
    \caption{Example of simulated Shack-Hartmann turbulent and noisy images.
      (a) Full Shack-Hartmann long-exposure image. (b) Subaperture short-exposure image.  
      \label{fig:im_simu}
    }  
  \end{center}
\end{figure}

Typical SH noisy images obtained after propagation through von K\'arm\'an
turbulence are shown in Fig.~\ref{fig:im_simu}. The detector read-out noise is
$\sigma_{\mathrm{e}^-}= 1$~e$^-$/pixel. More details on the image simulation
process are given in~\cite{voyez-p-12}.

Slopes and intensity fluctuations are extracted from the turbulent images.
Slopes are measured using a center of gravity (COG) algorithm, in windows of
$8 \times 8$ pixels,
centered on the maximum of each star. We do not use smaller windows because
doing so introduced a bias on the $C_n^2$ profile
reconstruction. Here, we cannot use larger windows either to avoid mixing
the two stars' signals.
The intensities, from which we deduce the
intensity fluctuations, correspond to the sum of all pixel intensities
included in the windows. 

\begin{figure}[htbp]
  \begin{center}\leavevmode
    \subfloat[]{
      \includegraphics[width=0.32\linewidth]{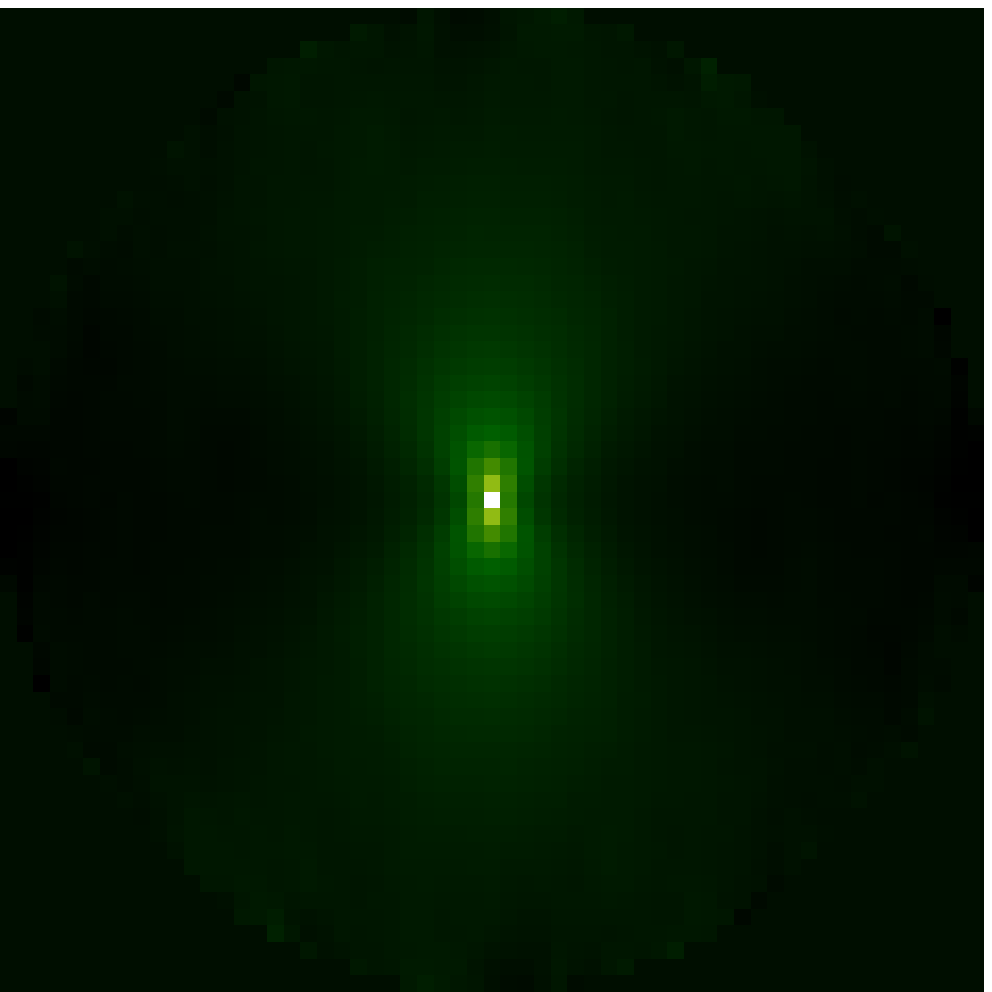}}
    \subfloat[]{
      \includegraphics[width=0.32\linewidth]{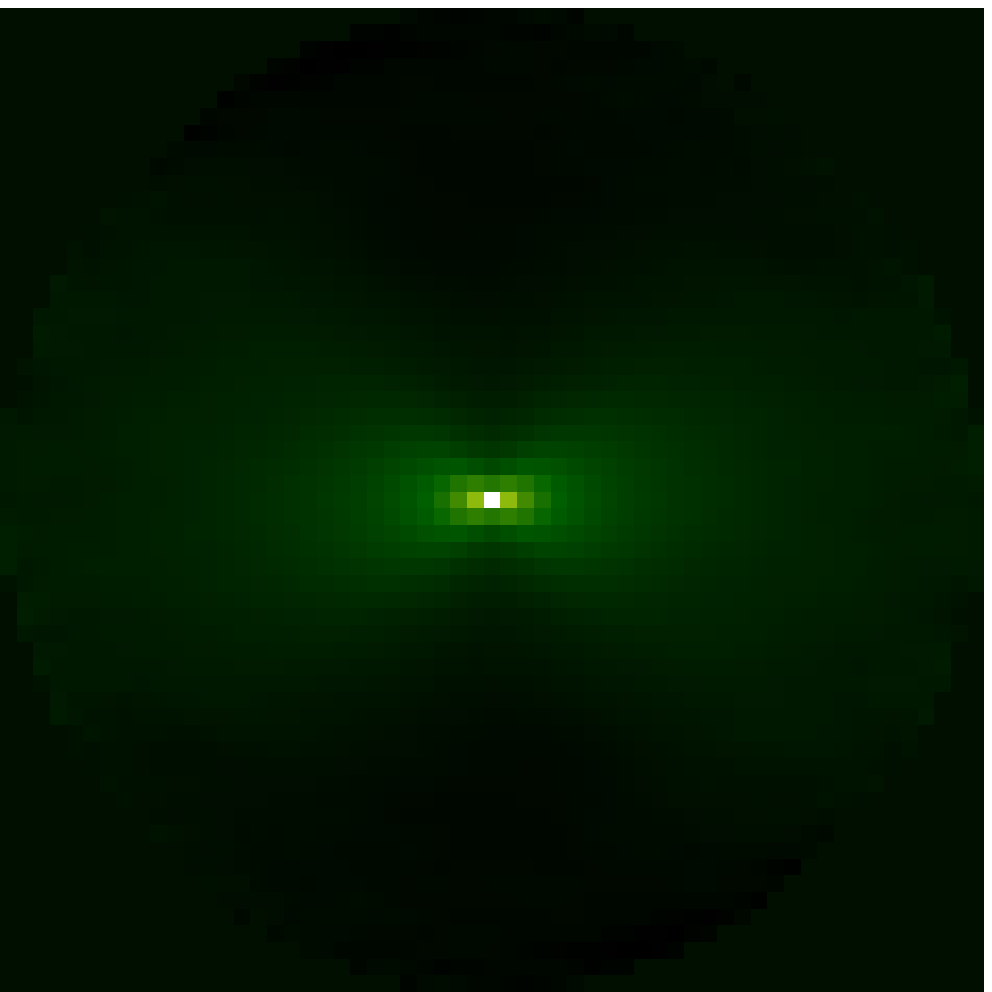}}
    \subfloat[]{
      \includegraphics[width=0.32\linewidth]{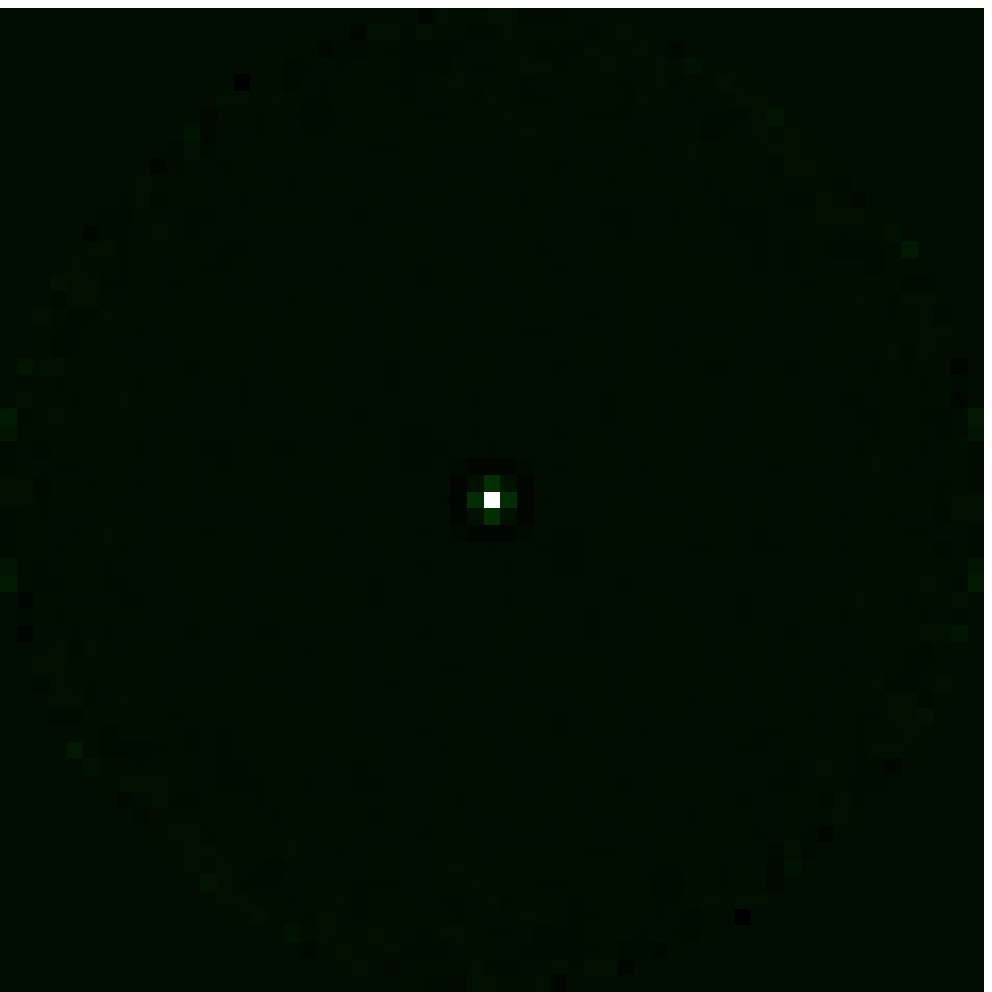}} \\
    \subfloat[]{
      \includegraphics[width=0.32\linewidth]{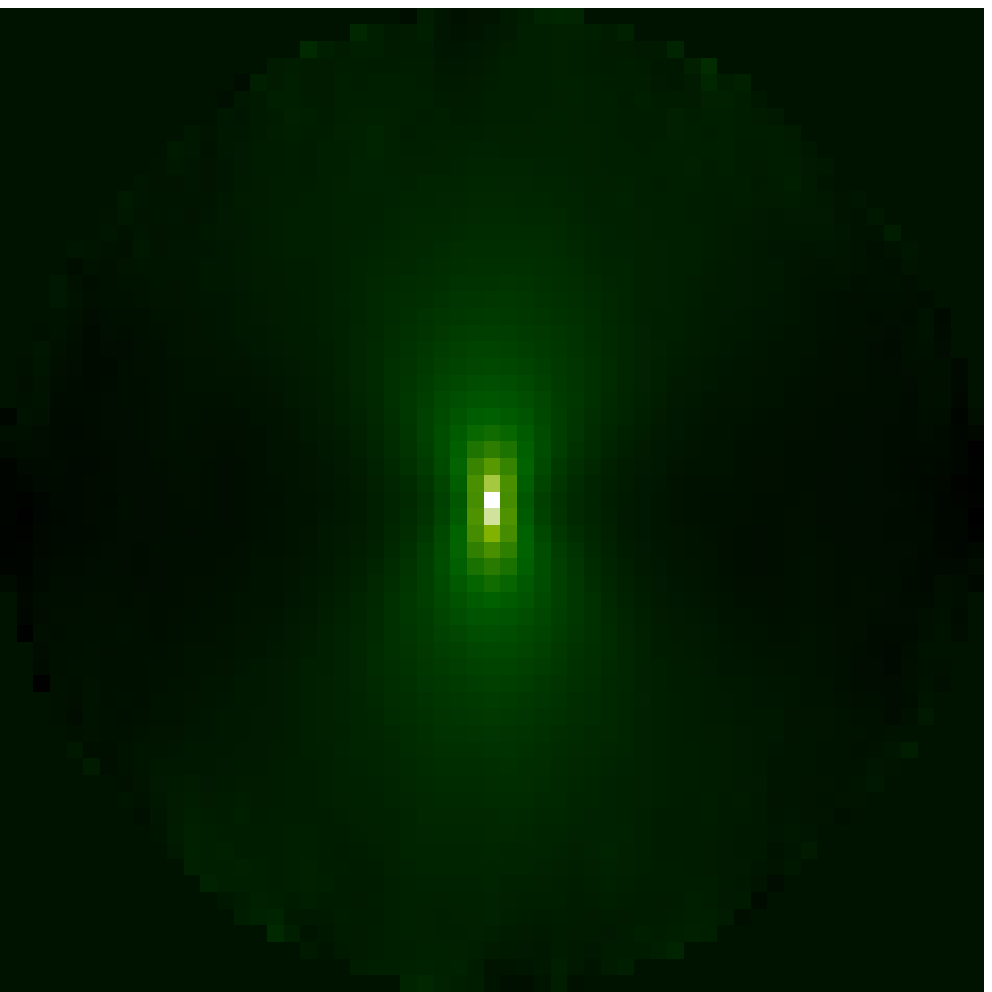}}
    \subfloat[]{
      \includegraphics[width=0.32\linewidth]{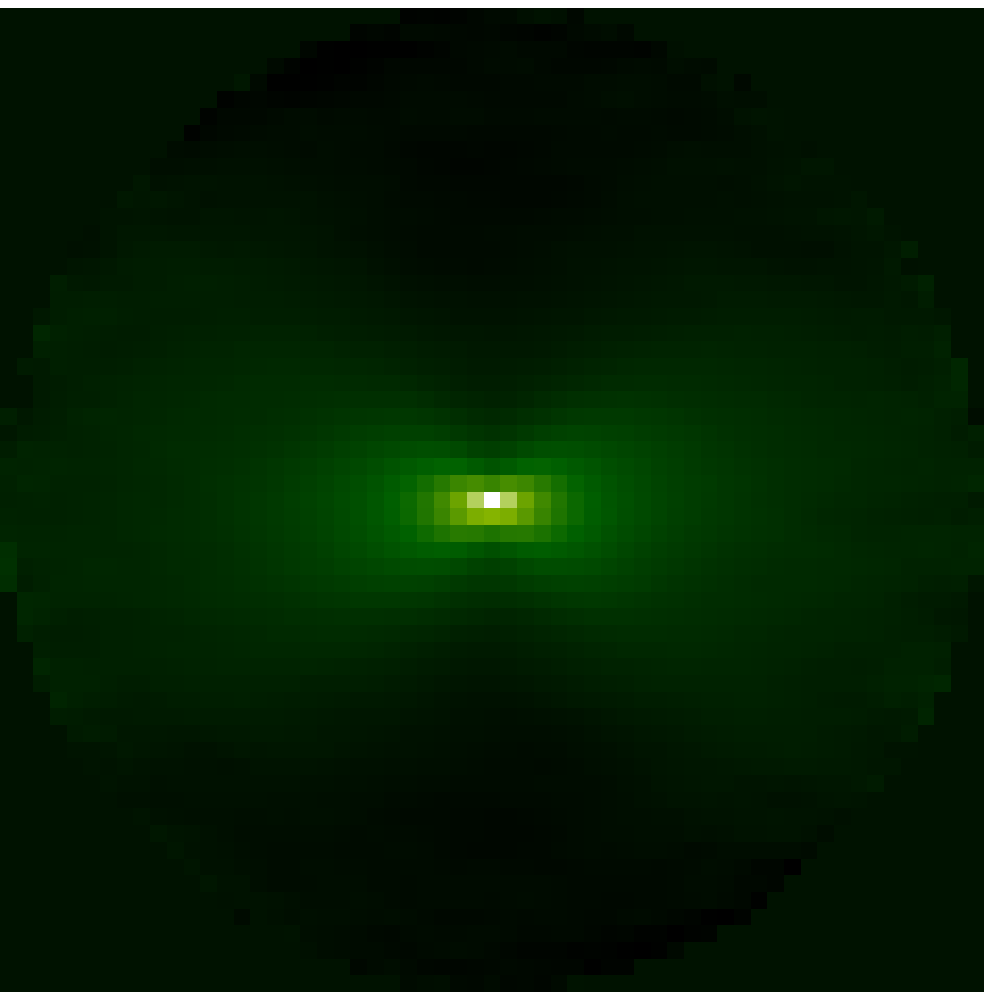}}
    \subfloat[]{
      \includegraphics[width=0.32\linewidth]{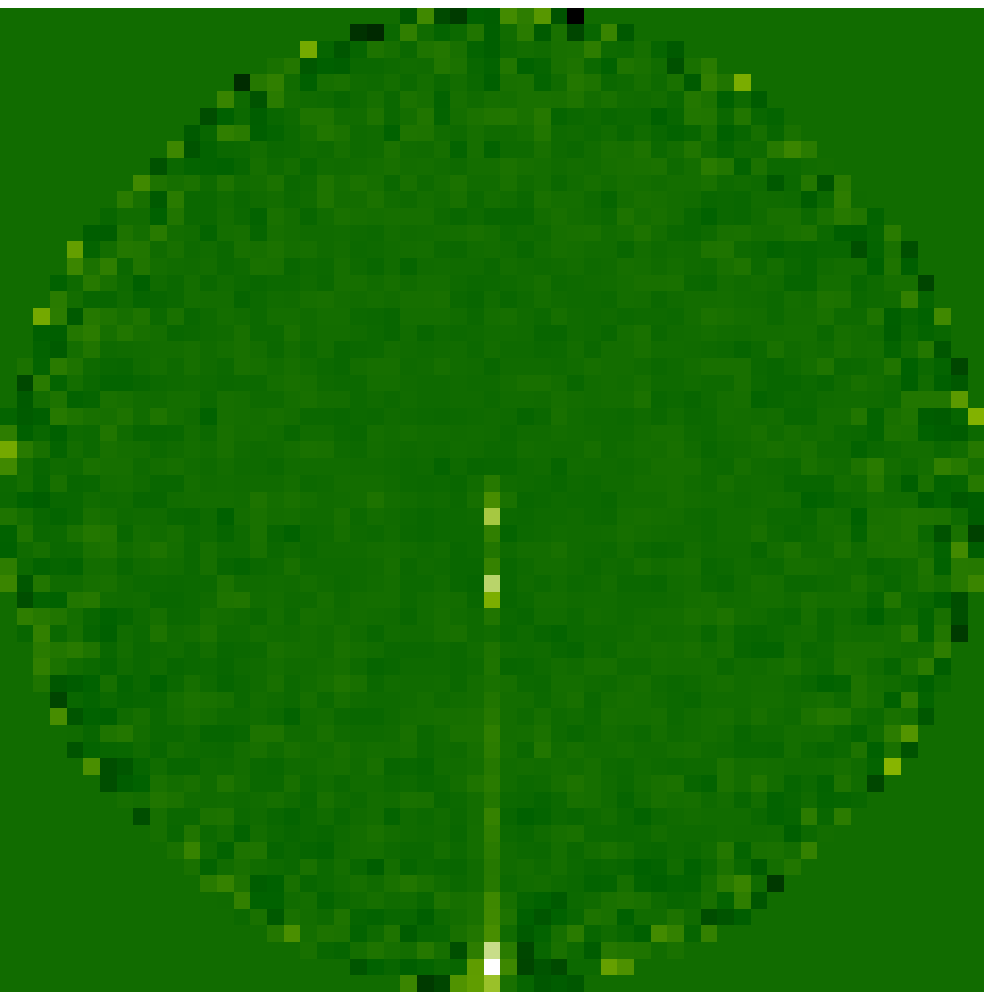}}
    \caption{Correlation maps from simulated slope and scintillation data. One pixel of these maps shows the correlation averaged over
      all pairs of subapertures with given separation. (a), (b), (c) Auto-correlation
      maps. (d), (e), (f) Cross-correlation maps. (a), (d) Correlations of
      $x$-slopes. (b), (e) Correlations of $y$-slopes. (c), (f) Correlations of scintillation.
      \label{fig:corr_simu}
    }  
  \end{center}
\end{figure}

Then we build correlation maps, presented in Fig.~\ref{fig:corr_simu}.
These maps show the correlation averaged over all pairs of subapertures with
given separation.
The auto-correlation maps have a maximum at their center.
They represent the response of the system to the integral of turbulence. We
can notice that the scintillation response is very narrow compared to the
slope response.
In the cross-correlation maps, the peak of correlation associated to the
turbulent layer at altitude $h$ is centered on $\mathrm{\pmb{\rho}}=
\mathrm{\pmb{\theta}}h$. In the cross-correlation map of scintillation, the peaks of correlation
associated to the turbulent layers at different altitudes are visible
in the bottom part of the map, in the alignment direction of the stars.
In the cross-correlation maps of slopes, only the peak of correlation
corresponding to $h=0$ is visible, at the center of the map. The peaks of
correlation associated to the other layers are also located at $\mathrm{\pmb{\theta}}h$, but because
of the width of the response and its decreasing strength with altitude, they
are not visible to the naked eye.

In CO-SLIDAR, we use both slope and scintillation responses to be sensitive to
low and high altitude turbulent layers.

\subsection{Reconstruction of the $C_n^2$ profiles}\label{sec:profiles_simu}
We now use the ML solution to retrieve the $C_n^2$ profile. Here, $\delta h
\simeq 500$~m, and $H_{\mathrm{max}} \simeq 15$~km, according to
Eqs.~(\ref{eq:res_slodar}) and~(\ref{eq:alt_max}). The detection noise bias is estimated
jointly with the $C_n^2$ profile and subtracted. We estimate $32$ layers, with
the same altitude sampling as for the theoretical $C_n^2$ input profile. 
\begin{figure}[htbp]
  \begin{center}\leavevmode
    \subfloat[]{
      \includegraphics[width=0.7\linewidth]{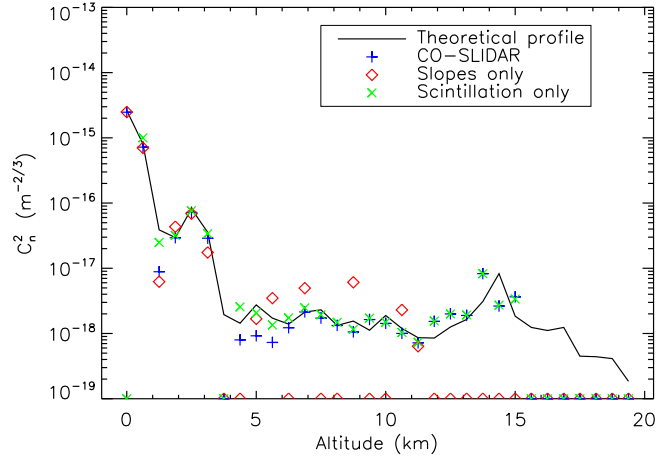}}\\
    \subfloat[]{
      \includegraphics[width=0.7\linewidth]{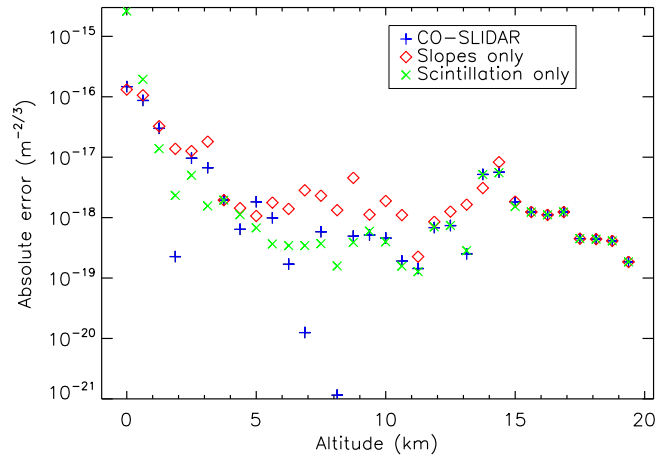}}
    \caption{(a) ML reconstruction of the $C_n^2$ profile from correlations of
      slopes only, of scintillation only and with the CO-SLIDAR method, in simulation. In all
      cases, the detection noise bias has been estimated jointly with the
      $C_n^2$ profile. (b) Absolute error on the reconstruction.
      \label{fig:profiles_simu}
    }
  \end{center}
\end{figure}
The results presented in Fig.~\ref{fig:profiles_simu} are compared to profiles
reconstructed from correlations of slopes only or of scintillation only.
Because of the positivity constraint, some values are estimated to zero by the
estimator. They are arbitrarily set to $10^{-19}$ m$^{-\nicefrac{2}{3}}$ for
the display. We also plot the absolute error on the reconstruction denoted
$|\left(C_{n_{\mathrm{est}}}^2\left(h_i\right)-C_{n_{\mathrm{th}}}^2\left(h_i\right)\right)|$.
$C_{n_{\mathrm{est}}}^2\left(h_i\right)$ is the estimated $C_n^2$ at altitude
$h_i$ ($i \in \left[1 \; ; \; 32\right] $) and
$C_{n_{\mathrm{th}}}^2\left(h_i\right)$ is the true $C_n^2$ at altitude $h_i$.
In this figure, we see that correlations of slopes allow a good reconstruction
of ground and low altitude layers, but provide a poor resolution at high
altitude (alternation of strong and zero values) and no sensitivity at all
above $12$~km. Conversely, correlations of scintillation lead to a good
reconstruction at low and high altitude, but they do not allow the estimation
of the ground layer.

CO-SLIDAR takes advantage of both correlations of slopes and of scintillation,
leading to a more complete reconstruction of low and high
altitude layers. CO-SLIDAR, as well as the other methods, does not succeed in estimating layers over $15$~km,
altitude which corresponds to $H_{\mathrm{max}}$ in this simulation. The Fried parameter estimated from the
CO-SLIDAR $C_n^2$ profile is $r_0 \simeq 5.8$~cm and the estimated variance of
log-amplitude is $\sigma_{\chi}^2 \simeq 0.020$. These values are very close
to the true ones: $r_{0_{\mathrm{th}}} \simeq 5.5$~cm, $\sigma_{\chi_{\mathrm{th}}}^2 \simeq
0.023$.

In Table~\ref{tab:rmse}, we present the Root Mean Square Error (RMSE), for
each reconstruction of the $C_n^2$ profile and for different altitude slices, defined
as:
\begin{equation}
RMSE =\left(\frac{1}{N_l}\sum_{i=1}^{N_l} \left(C_{n_{\mathrm{est}}}^2\left(h_i\right)-C_{n_{\mathrm{th}}}^2\left(h_i
    \right)\right)^2\right)^{\nicefrac{1}{2}},
\end{equation}
where $N_l$ is the number of layers considered. We divided the profile into
three altitude slices: the first two layers, the next four layers, and the
remaining high altitude layers. As the reconstruction from correlations of scintillation
only is not sensitive to the ground layer, we only consider the second layer
to calculate the RMSE in the first altitude slice.
\begin{table}[htbp]
\begin{center}\leavevmode
\begin{tabular}{|c|c|c|c|}
\cline{2-4}
\multicolumn{1}{c|}{} & \multirow{2}{*}{\vspace{5mm}First two layers} & \multicolumn{1}{c|}{\multirow{2}{*}{Four next layers}} & \multicolumn{1}{c|}{\multirow{2}{*}{High altitude layers}} \\
\multicolumn{1}{c|}{} & (Second layer) & & \\
\hline
\multicolumn{1}{|c|}{\multirow{2}{*}{Slopes only (m$^{-\nicefrac{2}{3}}$)}} &
\multirow{2}{*}{\vspace{5mm}$1.2 \times 10^{-16}$} & \multicolumn{1}{c|}{\multirow{2}{*}{$2.1 \times 10^{-17}$}} & \multicolumn{1}{c|}{\multirow{2}{*}{$2.4 \times 10^{-18}$}} \\
\multicolumn{1}{|c|}{} & $\left(1.1 \times 10^{-16}\right)$ & & \\
\hline
\multicolumn{1}{|c|}{\multirow{2}{*}{Scintillation only (m$^{-\nicefrac{2}{3}}$)}} &
\multirow{2}{*}{\vspace{5mm} not applicable} & \multicolumn{1}{c|}{\multirow{2}{*}{$7.5 \times 10^{-18}$}} & \multicolumn{1}{c|}{\multirow{2}{*}{$1.7 \times 10^{-18}$}} \\
\multicolumn{1}{|c|}{} & $\left(1.9 \times 10^{-16}\right)$ & & \\
\hline
\multicolumn{1}{|c|}{\multirow{2}{*}{CO-SLIDAR (m$^{-\nicefrac{2}{3}}$)}} &
\multirow{2}{*}{\vspace{5mm}$1.2\times 10^{-16}$} & \multicolumn{1}{c|}{\multirow{2}{*}{$1.6 \times 10^{-17}$}} & \multicolumn{1}{c|}{\multirow{2}{*}{$1.7 \times 10^{-18}$}} \\
\multicolumn{1}{|c|}{} & $\left(8.7 \times 10^{-17}\right)$ & & \\
\hline
\end{tabular}
\caption{RMSE in function of the altitude slice for the ML reconstruction of
  the $C_n^2$ profile, from correlations of slopes only, of scintillation only, and
  with the CO-SLIDAR method.\label{tab:rmse}}
\end{center}
\end{table}

The results in Table ~\ref{tab:rmse} show that for the first two layers, the
CO-SLIDAR reconstruction is similar to the reconstruction from correlations of
slopes and better than the reconstruction from correlations of scintillation.
For high altitude layers, the CO-SLIDAR reconstruction is similar to the
reconstruction from correlations of scintillation and better than the
reconstruction from correlations of slopes. For the four layers between $1.25$
and $3.125$~km, the CO-SLIDAR reconstruction is better than the reconstruction
from correlations of slopes, but is surprisingly worse than the reconstruction
from correlations of scintillation. In principle, using  more data should not
decrease the quality of the reconstruction. This effect is possibly due to the
fact that we \emph{estimate} the covariance matrix of the convergence
noise $C_{\mathrm{conv}}$ using the pseudo-measurements. It should be investigated in more
details in a future work.

\begin{figure}[htbp]
  \begin{center}\leavevmode
    \includegraphics[width=0.7\linewidth]{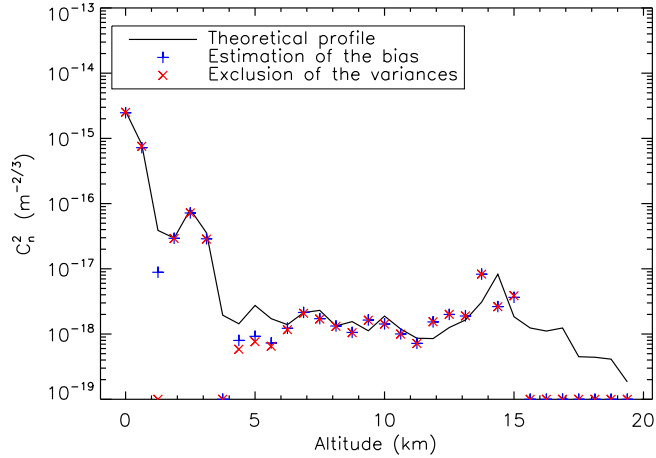}
    \caption{ML CO-SLIDAR reconstruction of the $C_n^2$ profile, with joint
      estimation of the detection noise bias and with exclusion of the variances
      from the direct problem, in simulation.
      \label{fig:profiles_noise_comp_simu}
    }
  \end{center}
\end{figure}

The two methods to subtract the detection noise bias are now compared in
Fig.~\ref{fig:profiles_noise_comp_simu} for the CO-SLIDAR reconstruction. 
In one case, as previously explained, we estimate the detection noise bias jointly with the
turbulence profile, whereas in the other case we simply exclude the noisy
pseudo-measurements, \textit{i.e.} the variances of slopes and of scintillation,
from the direct problem.

In the joint estimation, the variances of noises that we estimated were close
to the expected ones $(\sigma_{s^x_{\mathrm{n\text{-}th}}}^{2}=0.26$
rad$^{2}$, $\sigma_{s^y_{\mathrm{n\text{-}th}}}^{2}=0.28$ rad$^{2}$,
$\sigma_{\delta i_{\mathrm{n\text{-}th}}}^{2}=8.7 \times 10^{-3})$, but
slightly higher $(\sigma_{s^x_{\mathrm{n\text{-}est}}}^{2}=2.8$ rad$^{2}$,
$\sigma_{s^y_{\mathrm{n\text{-}est}}}^{2}=0.32$ rad$^{2}$, $\sigma_{\delta
  i_{\mathrm{n\text{-}est}}}^{2}=1.2\times 10^{-2})$. By subtracting the expected values from the estimated ones,
we found a residual term. We noted that this term was identical to the one
retrieved by the estimator when
performing joint estimation with data free from detection noises
$(\sigma_{s^x_{\mathrm{n\text{-}res}}}^{2}=3,7 \times 10^{-2}$
rad$^{2}$, $\sigma_{s^y_{\mathrm{n\text{-}res}}}^{2}=5,7 \times 10^{-2}$
rad$^{2}$, $\sigma_{\delta i_{\mathrm{n\text{-}res}}}^{2}=3,1 \times 10^{-3})$. 
Moreover, we noticed in our simulations that the reconstruction of the
$C_n^2$ profile from data free from detection noise was more accurate when
estimating the bias. Therefore this small residual term can be identified as
the one corresponding to model errors (for example, in the simulation, the error due to the size
of the window for the measurements). Thus, the joint estimation method
allows one to estimate accurately the $C_n^2$ profile and the variances of noises, and
is capable of estimating biases other than those due to detection noises, adding
extra degrees of freedom to the estimation.  

In the second method, we discard the variances of slopes and of scintillation from the pseudo-measurements.
When we compare the two methods, we find that they give very similar
results and allow a good reconstruction of the turbulent profile, so we have
two efficient methods to subtract the bias due to detection noises.

\begin{figure}[htbp]
  \begin{center}\leavevmode
    \includegraphics[width=0.7\linewidth]{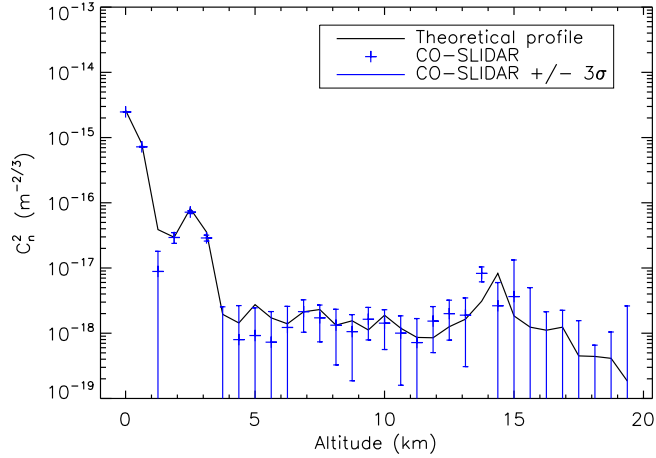}
    \caption{ML CO-SLIDAR reconstruction of the $C_n^2$ profile, with joint estimation of the detection noise
      bias, and with the estimated $3\sigma$ error bars, in simulation.
      \label{fig:profile_error_bars_simu}
    }
  \end{center}
\end{figure}

Then we compute the $3\sigma$ error bars, and we represent them together with
the estimated profile, as shown in Fig.~\ref{fig:profile_error_bars_simu}.
The $C_n^2$ profile is retrieved with the joint estimation method.
We note that the relative error, defined as the absolute error divided
by the $C_n^2$ value at a given altitude, is weaker for turbulent layers with
strong $C_n^2$ values and low altitudes.
Indeed, these layers correspond to a
better signal to noise ratio (SNR). Moreover, cross-correlations corresponding
to these altitudes are estimated with a larger number of samples, because they
are associated to more pairs of subapertures.
Finally, we can see that the true $C_n^2$ profile is well included within
these $3\sigma$ error bars, excepted the layers located at $1.25$ and $13.75$~km.

\begin{figure}[htbp]
  \begin{center}\leavevmode
    \subfloat[]{
      \includegraphics[width=0.7\linewidth]{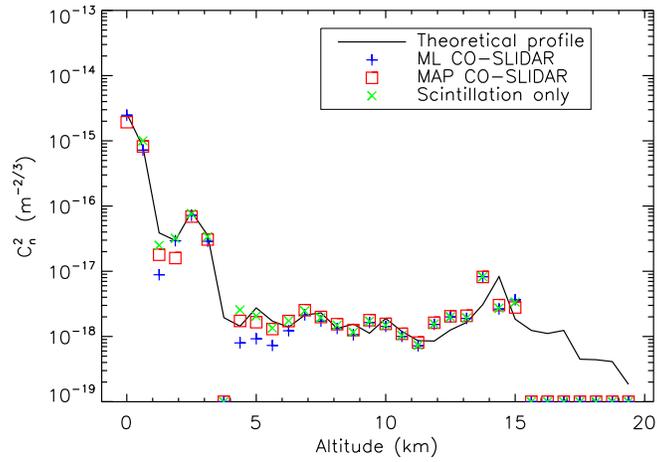}}\\
    \subfloat[]{
      \includegraphics[width=0.7\linewidth]{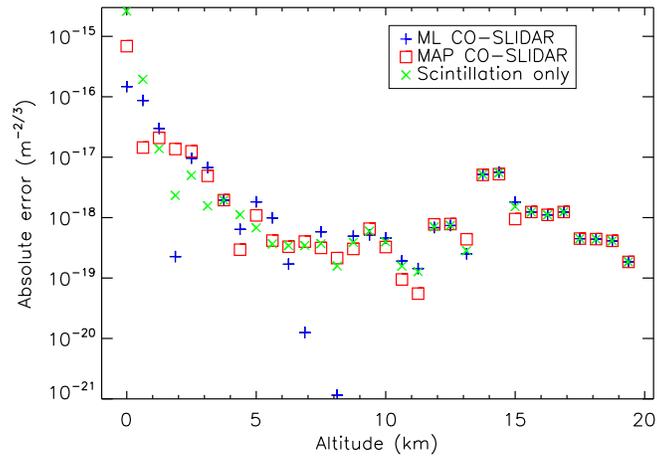}}
    \caption{(a) ML and MAP CO-SLIDAR reconstructions of the $C_n^2$ profile and comparison with the reconstruction from correlations
      of scintillation only, in simulation. (b) Absolute error on the reconstruction.
      \label{fig:profiles_ml_map_simu}
    }
  \end{center}
\end{figure}

In Fig.~\ref{fig:profiles_ml_map_simu}, we finally compare the ML CO-SLIDAR
reconstruction of the $C_n^2$ profile with the reconstruction from correlations of
scintillation only and with the MAP CO-SLIDAR reconstruction for a weakly
regularized case. 
The regularization parameter $\beta$ has been chosen so that the likelihood
and the prior terms of Eq.~\ref{eq:MAP_J} have approximately the same order of
magnitude. We have increased it until the estimation of the ground layer,
on which there is a very good SNR, begins to be biaised towards zero. Note that
this tuning strategy is also applicable to experimental data.
We also plot the absolute error on the reconstruction.
In this figure, we see that the MAP $C_n^2$ profile is smoother than the ML
profile as expected. We note slight differences between the ML and the MAP
profiles but the two approaches allow a good reconstruction, as quantified by
Table~\ref{tab:rmse_map}.

In Table~\ref{tab:rmse_map}, we present the Root Mean Square Error (RMSE) for 
the ML and MAP CO-SLIDAR reconstructions and for the reconstruction from
correlations of scintillation only.
The results in Table ~\ref{tab:rmse_map} show that for the first two layers,
the MAP reconstruction is still better than the reconstruction from
correlations of scintillation only, but worse than the ML reconstruction. This is
due to this kind of regularization, where the constraint applies mainly on
strong turbulent layers. In contrast, the reconstruction of high altitude
layers is slightly better with the MAP solution than with the ML solution and
than the reconstruction from correlations of scintillation only.
For the layers between $1.25$ and $3.125$~km, the reconstruction from
correlations of scintillation is still better, probably for the reasons
previously mentioned. 
\begin{table}[htbp]
\begin{center}\leavevmode
\begin{tabular}{|c|c|c|c|}
\cline{2-4}
\multicolumn{1}{c|}{} & \multirow{2}{*}{\vspace{5mm}First two layers} & \multicolumn{1}{c|}{\multirow{2}{*}{Four next layers}} & \multicolumn{1}{c|}{\multirow{2}{*}{High altitude layers}} \\
\multicolumn{1}{c|}{} & (Second layer) & & \\
\hline
\multicolumn{1}{|c|}{\multirow{2}{*}{Scintillation only (m$^{-\nicefrac{2}{3}}$)}} &
\multirow{2}{*}{\vspace{5mm}not applicable} & \multicolumn{1}{c|}{\multirow{2}{*}{$7.5 \times 10^{-18}$}} & \multicolumn{1}{c|}{\multirow{2}{*}{$1.7 \times 10^{-18}$}} \\
\multicolumn{1}{|c|}{} & $\left(1.9 \times 10^{-16}\right)$ & & \\
\hline
\multicolumn{1}{|c|}{\multirow{2}{*}{ML CO-SLIDAR (m$^{-\nicefrac{2}{3}}$)}} &
\multirow{2}{*}{\vspace{5mm}$1.2\times 10^{-16}$} & \multicolumn{1}{c|}{\multirow{2}{*}{$1.6 \times 10^{-17}$}} & \multicolumn{1}{c|}{\multirow{2}{*}{$1.7 \times 10^{-18}$}} \\
\multicolumn{1}{|c|}{} & $\left(8.7 \times 10^{-17}\right)$ & & \\
\hline
\multicolumn{1}{|c|}{\multirow{2}{*}{MAP CO-SLIDAR (m$^{-\nicefrac{2}{3}}$)}} &
\multirow{2}{*}{\vspace{5mm}$4.9\times 10^{-16}$} & \multicolumn{1}{c|}{\multirow{2}{*}{$1.4 \times 10^{-17}$}} & \multicolumn{1}{c|}{\multirow{2}{*}{$1.6 \times 10^{-18}$}} \\
\multicolumn{1}{|c|}{} & $\left(1.4 \times 10^{-17}\right)$ & & \\
\hline
\end{tabular}
\caption{RMSE in function of the altitude slice, for the reconstruction of the
  $C_n^2$ profile from correlations of scintillation only and for the ML and MAP CO-SLIDAR reconstructions.\label{tab:rmse_map}}
\end{center}
\end{table}
To improve the accuracy of the MAP estimation, we could
use an adaptive regularization~\cite{sauvage-p-06}, designed to take
into account the dynamic range of the $C_n^2$ profile.  

We have shown in this section that CO-SLIDAR gives good results in
simulation, and allows one to estimate the $C_n^2$ profile over the whole range
of altitudes, with a sub-kilometric resolution. The next step is to examine its on-sky
performance.

\section{On-sky results}\label{sec:on_sky_results}
In this section we describe the on-sky experiment and present the very first
on-sky results of the CO-SLIDAR instrument. In Subsection~\ref{sec:instrument}
we detail the CO-SLIDAR instrument. Subsection~\ref{sec:data_exp} is dedicated to observations and data analysis. The
estimated $C_n^2$ profiles are shown in Subsection~\ref{sec:profiles_exp}.
These CO-SLIDAR profiles are compared to $C_n^2$ profiles deduced from
NCEP/NCAR Reanalysis in Subsection~\ref{sec:COSLIDAR_NCEPNCAR}.  

\subsection{The CO-SLIDAR instrument}\label{sec:instrument}
The experiment took place on the Plateau de Calern, at the Observatoire de la
Côte d'Azur, near Nice, France. We used the $1.5$~m MeO telescope,
with a central obscuration of $30$~\%, coupled to a $30\times30$ subaperture
SH, hence the subaperture diameter is $d_{\mathrm{sub}}=5$~cm. The CO-SLIDAR
instrument was mounted behind the Coudé train of the telescope, composed of
seven mirrors and one doublet. Two lenses of focal lengths $800$~mm and $56$~mm were used to image the telescope pupil onto the $30\times 30$ lenslet
array. The microlens array had a pitch of $143$ $\mathrm{\mu}$m and a focal length of
$3.6$~mm. As it was too short to form an image directly on the detector array,
we used an additional pair of lenses of focal lengths $120$~mm and $150$~mm to transfer the focal plane. The
observation wavelength was $\lambda=517$~nm, with $\Delta \lambda = 96$~nm. The
camera used was an Andor-iXon3-885 electron
multiplication CCD (EMCCD) with a quantum efficiency of about $50$~\%, and a
detector read-out noise close to one e$^-$/pixel. The final plate scale on the
SH is $1.2$ arcsec/pixel.

\subsection{Observations and data analysis}\label{sec:data_exp}
\begin{figure}[htbp]
  \begin{center}\leavevmode
    \subfloat[]{
      \includegraphics[width=0.4\linewidth]{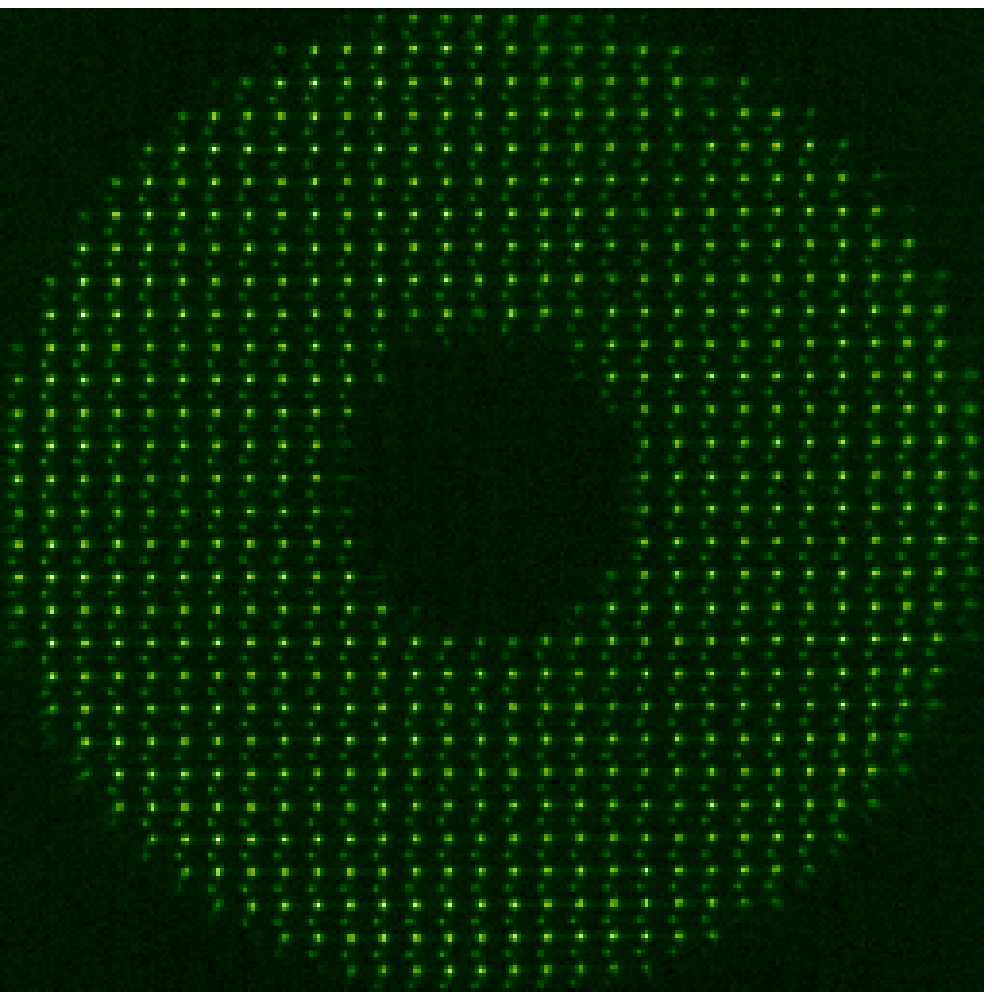}}
    \subfloat[]{
      \includegraphics[width=0.4\linewidth]{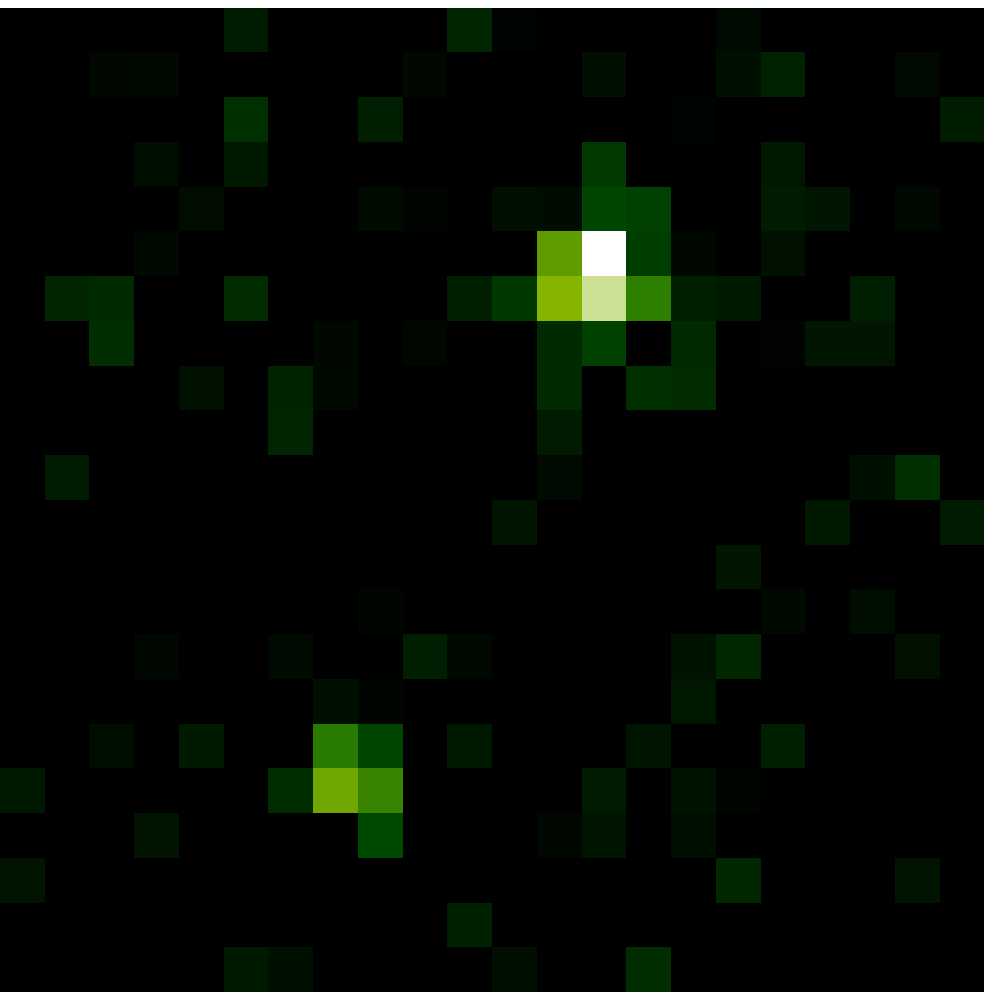}}
    \caption{Experimental Shack-Hartmann turbulent images, for a $3$~ms
      exposure time. (a) Full Shack-Hartmann long-exposure image. (b)
      Subaperture short-exposure image.
      \label{fig:im_exp}
    }  
  \end{center}
\end{figure}

Observations were done on May 2012, on the double star Mizar AB. We selected the
data from May, $15^{\mathrm{th}}$, around $01 \!\! :\!\! 00$~UT. The zenith angle of the
binary star was $\zeta = 35\degree$. The exposure time was
$t_{\mathrm{exp}}=3$~ms, to freeze the turbulence. The separation between the
two components is $\theta=14.4$'' and their visible magnitudes are $2.23$ and
$3.88$, leading to about $260$ and $60$ photons per subaperture and per frame.
We recorded sequences of $1000$ images at $15$~Hz, so the sequence duration is
about $1$~min. Typical on-sky images are shown in Fig.~\ref{fig:im_exp}.

We extract slopes and scintillation from these images, in windows of
$9 \times 9$ pixels. We check the statistics of
turbulence with the two kinds of data. From slopes we compute the Zernike
coefficient variances, showing a Kolmogorov turbulence with an outer scale effect,
noticeable on the tip-tilt. We check the hypothesis of the weak
perturbation regime using intensities and intensity fluctuations, by fitting a
log-normal distribution and estimating $\sigma_{\chi_{\mathrm{data}}}^2$. The
latter was found to be far below $0.3$, validating the hypothesis of
weak perturbations.

\begin{figure}[htbp]
  \begin{center}\leavevmode
    \subfloat[]{
      \includegraphics[width=0.32\linewidth]{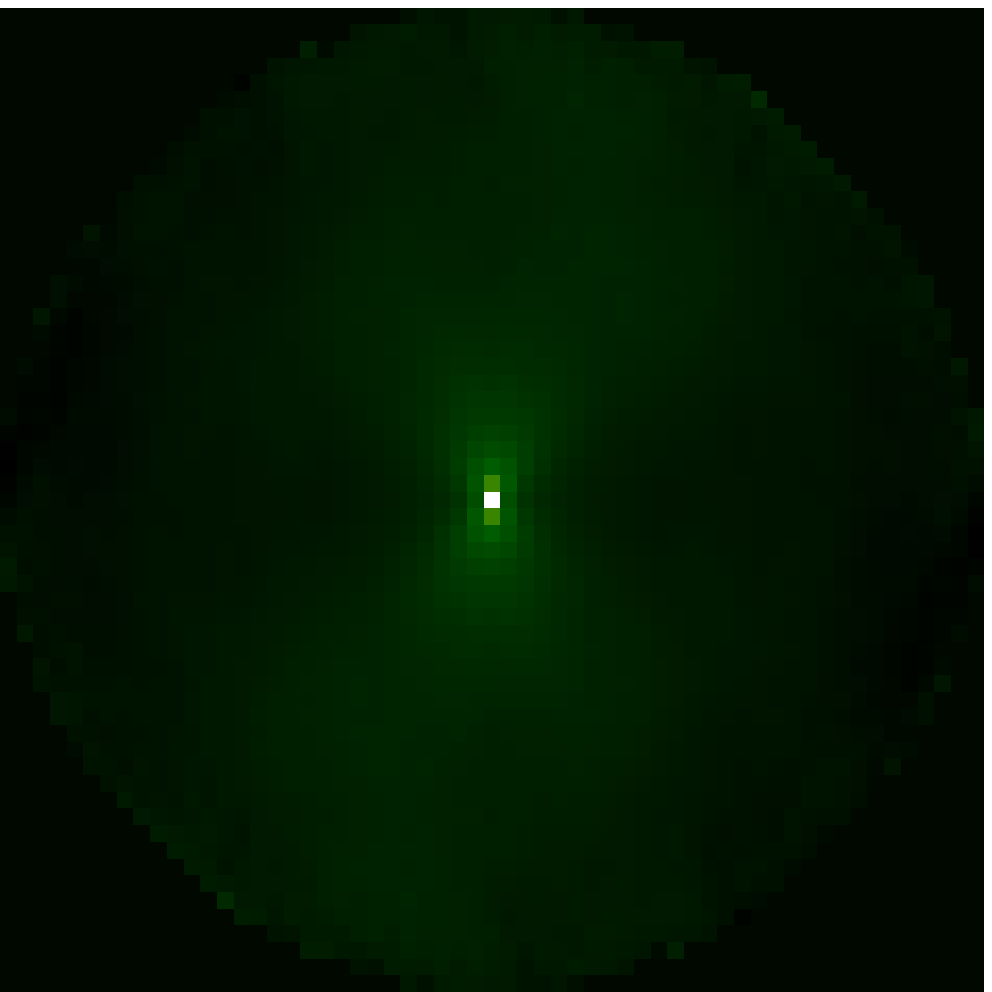}}
    \subfloat[]{
      \includegraphics[width=0.32\linewidth]{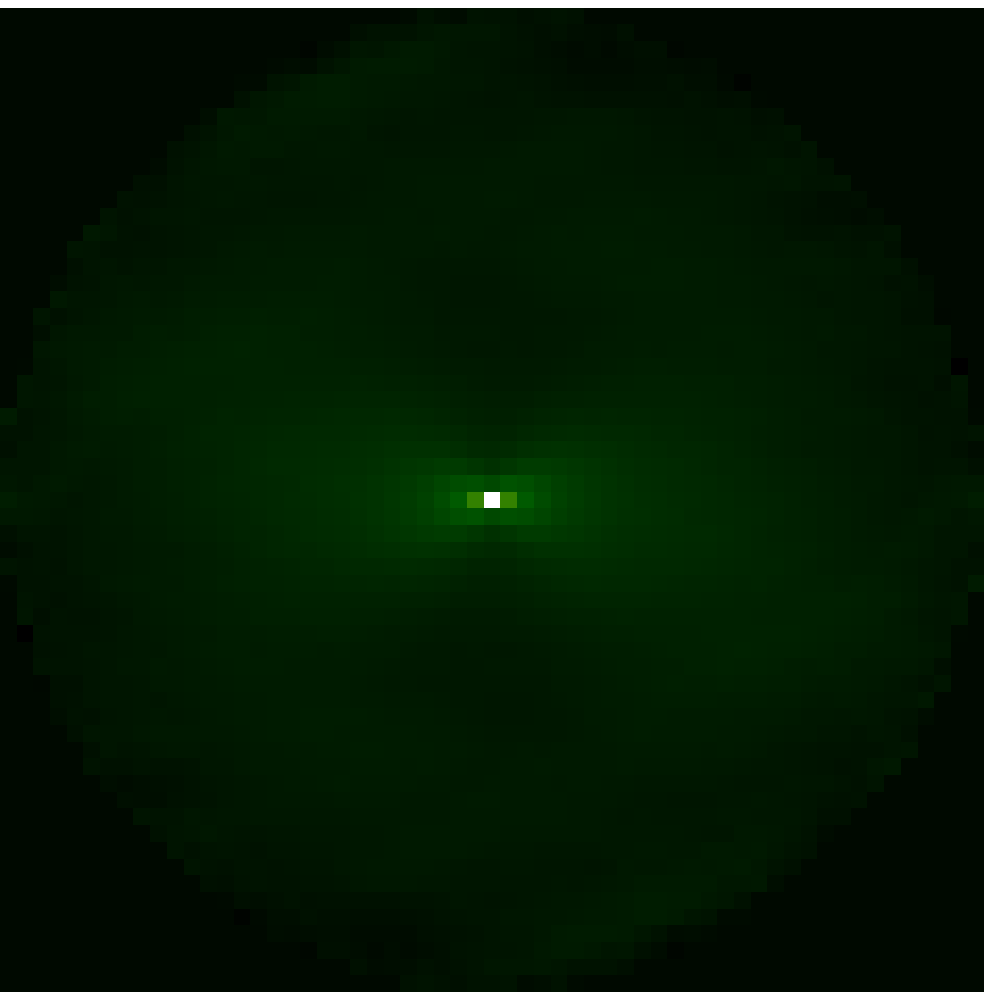}}
    \subfloat[]{
      \includegraphics[width=0.32\linewidth]{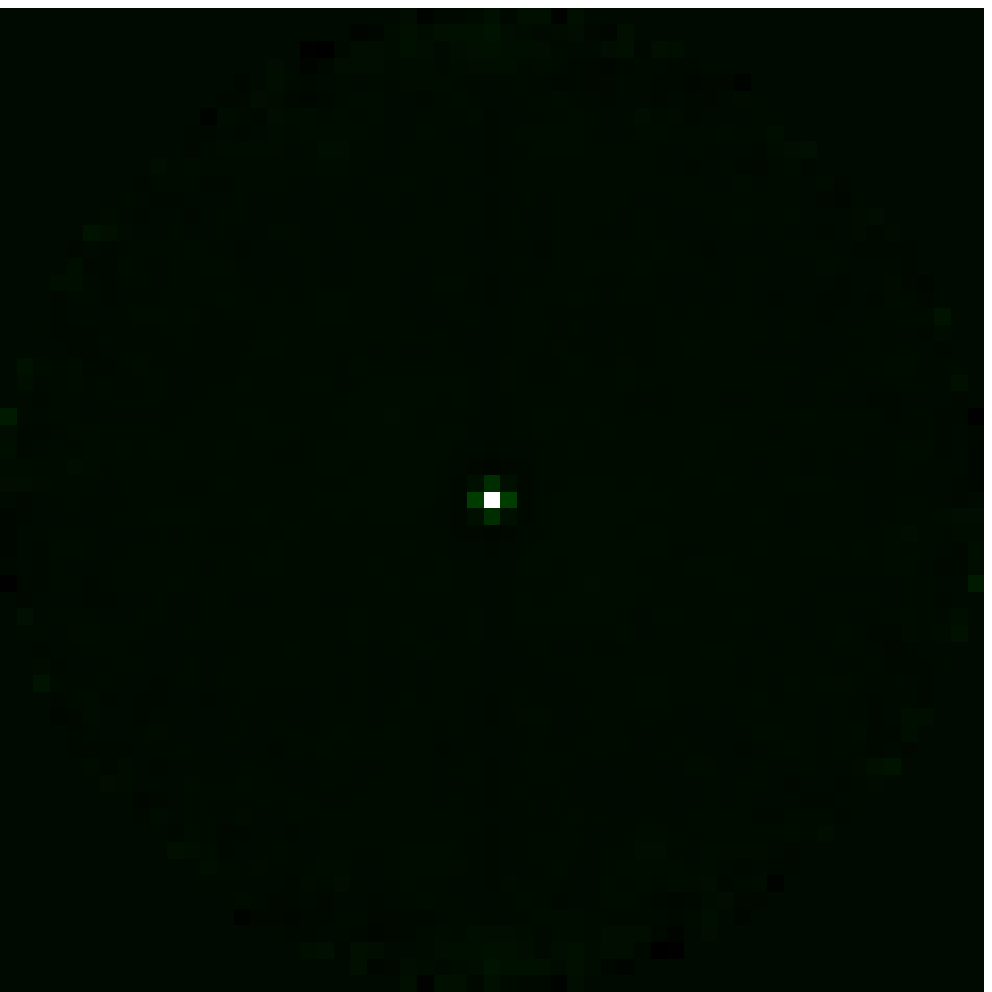}} \\
    \subfloat[]{ 
      \includegraphics[width=0.32\linewidth]{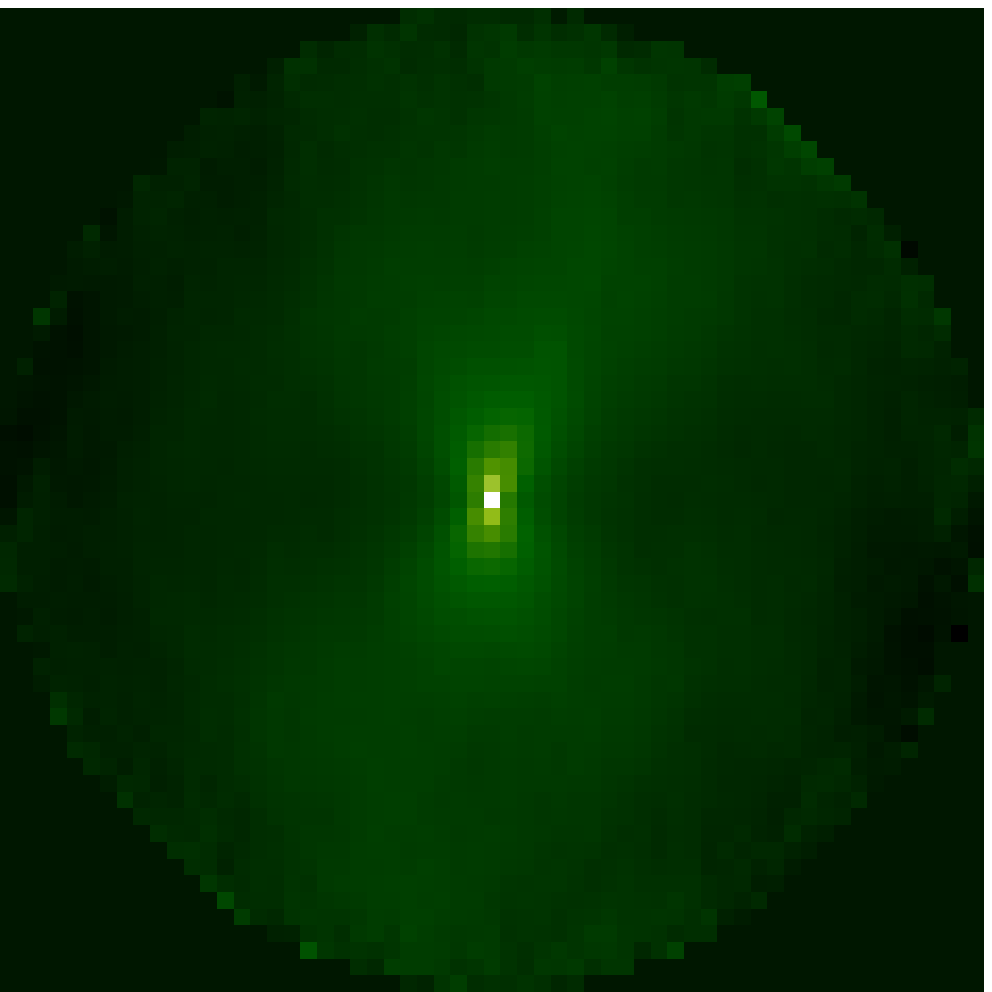}}
    \subfloat[]{
      \includegraphics[width=0.32\linewidth]{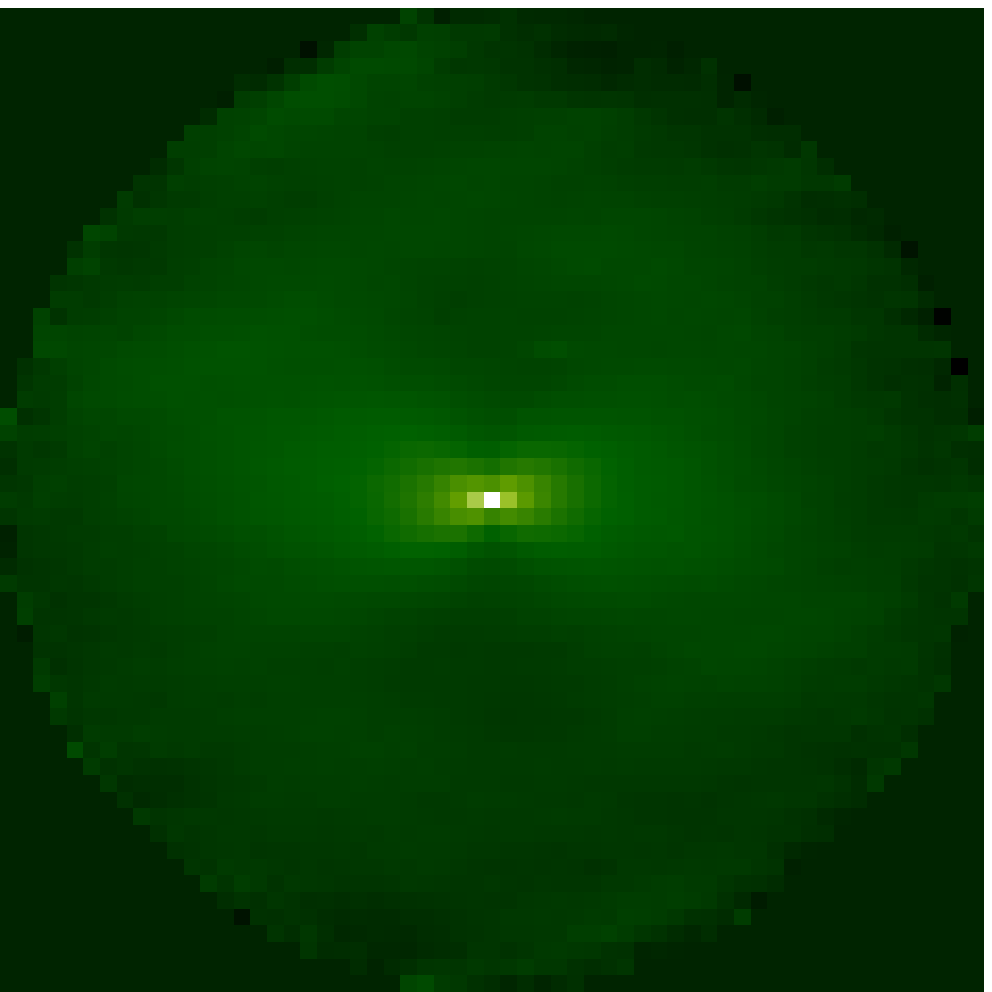}}
    \subfloat[]{
      \includegraphics[width=0.32\linewidth]{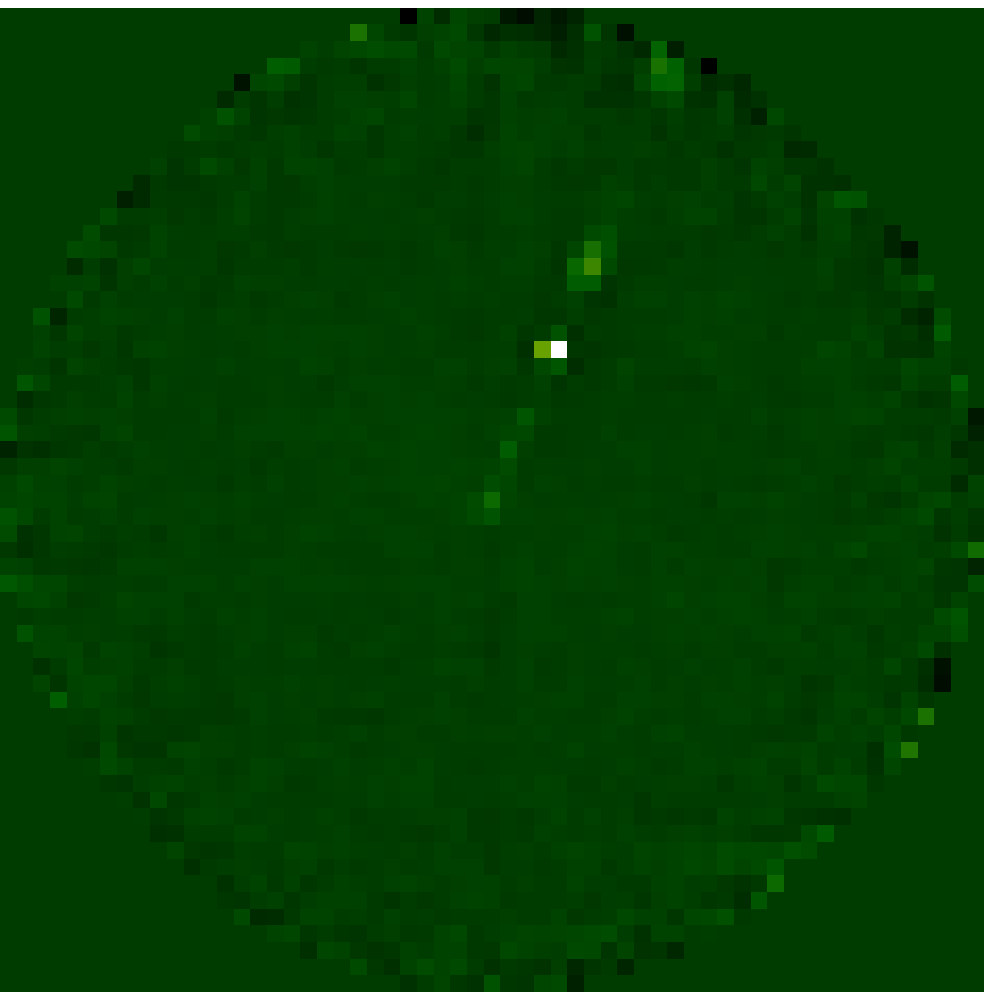}}
    \caption{Correlation maps from experimental slope and scintillation data.
      (a), (b), (c) Auto-correlation
      maps. (d), (e), (f) Cross-correlation maps. (a), (d) Correlations of
      $x$-slopes. (b), (e) Correlations of $y$-slopes. (c), (f) Correlations of scintillation.
      \label{fig:corr_exp}
    }  
  \end{center}
\end{figure}
The correlation maps in Fig.~\ref{fig:corr_exp} present a pattern similar to
the one obtained in the simulation. The cross-correlation map of scintillation
shows peaks of correlation in the top right quarter of the map, in the
alignment direction of the stars, representing the turbulent layers'
signatures. The comments on the other maps are the same as in the
simulation case.

These verifications confirm the data consistency with the model hypotheses and that we
can use the pseudo-measurements to estimate the $C_n^2$ profiles.

\subsection{Estimation of the turbulence profiles}\label{sec:profiles_exp}
As we use a von K\'arm\'an model for turbulence, we have to choose an outer
scale $L_0$. We assume that $L_0=27$~m, which is the median outer
scale observed at the Plateau de Calern~\cite{conan-t-00}. We checked that the
results were not significantly affected by the outer scale choice in the range
$\left[10 \; ; 50 \right]$~m and we found very similar $C_n^2$ profiles. We estimate $30$ layers. Here, $\delta h \simeq
600$~m and $H_{\mathrm{max}} \simeq 17$~km, using Eqs.~(\ref{eq:res_slodar})
and~(\ref{eq:alt_max}), corrected from the zenith angle $\zeta$.

\begin{figure}[htbp]
  \begin{center}\leavevmode
    \includegraphics[width=0.7\linewidth]{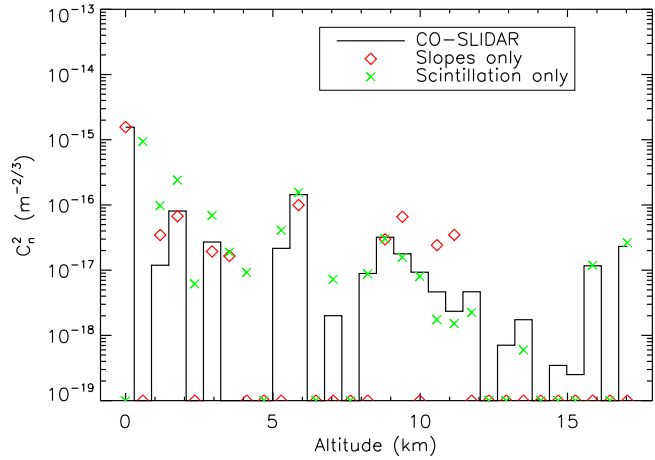} \\      
    \caption{ML reconstruction of the $C_n^2$ profile from correlations of
      slopes only, of scintillation only and with the CO-SLIDAR method.
      $1$~min dataset acquired around $01 \!\!:\!\!00$~UT, on May $15^{th}$,
      2012, with a $3$~ms exposure time. In all cases, the detection noise
      bias has been estimated jointly with the $C_n^2$ profile.
      \label{fig:profiles_exp}
    }  
  \end{center}
\end{figure}
The $C_n^2$ profiles are estimated with the ML solution, from correlations of
slopes only, of scintillation only and with the CO-SLIDAR method. For each
reconstruction, the detection noise bias is estimated jointly with the $C_n^2$
profile. The results are presented in Fig.~\ref{fig:profiles_exp}. At high
altitude, we find a good agreement between the CO-SLIDAR reconstruction and
that from correlations of scintillation. As in the simulated case, the
turbulence estimated from correlations of slopes is somewhat different,
alternation of stronger and zero values. At low altitude, we observe a good
agreement between the CO-SLIDAR reconstruction and that from correlations of
slopes, the turbulence estimated from correlations of scintillation being
stronger. However without any true profile, it is difficult to discard this
solution.

\begin{figure}[htbp]
  \begin{center}\leavevmode
    \includegraphics[width=0.7\linewidth]{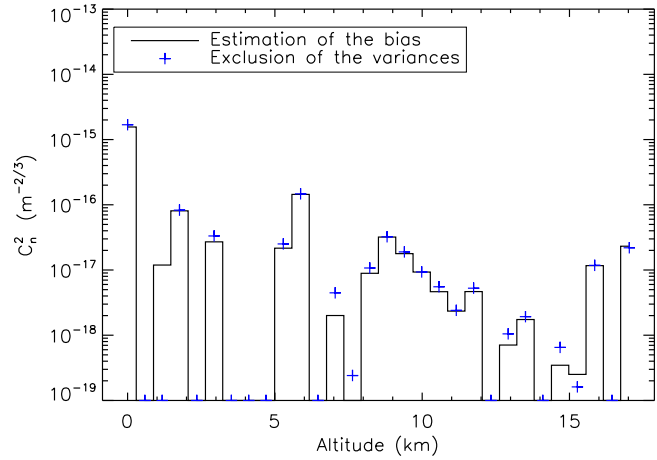} \\      
    \caption{ML CO-SLIDAR reconstruction of the $C_n^2$ profile, with joint
      estimation of the detection noise bias and with exclusion of the variances
      from the direct problem. $1$~min dataset from May $15^{th}$, 2012, around
      $01 \!\! : \!\! 00$~UT.
      \label{fig:profiles_noise_comp_exp}
    }  
  \end{center}
\end{figure}

In Fig.~\ref{fig:profiles_noise_comp_exp}, we compare the ML CO-SLIDAR $C_n^2$ profiles
reconstructed with joint estimation of the detection noise bias, and with
exclusion of the variances from the direct problem.  
We obtained very similar results with the two estimations. We note that
the $C_n^2$ values estimated when discarding the variances are slightly higher
than when we estimate the noise detection bias. In the following, we will
always use the joint estimation to reconstruct the $C_n^2$ profile.

\begin{figure}[htbp]
  \begin{center}\leavevmode
    \includegraphics[width=0.7\linewidth]{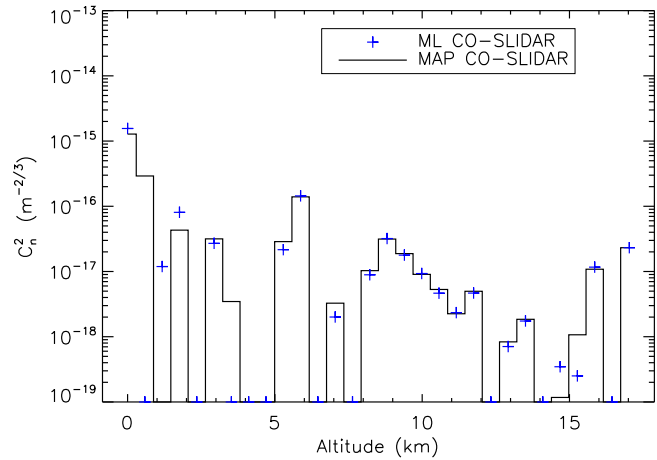}
    \caption{ML and MAP CO-SLIDAR reconstructions of the $C_n^2$ profile.
      $1$~min dataset from May $15^{th}$, 2012, around $01 \!\! : \!\! 00$~UT. 
      \label{fig:profiles_ml_map_exp}
    }  
  \end{center}
\end{figure}

\begin{figure}[htbp]
  \begin{center}\leavevmode
    \subfloat[]{
      \includegraphics[width=0.65\linewidth]{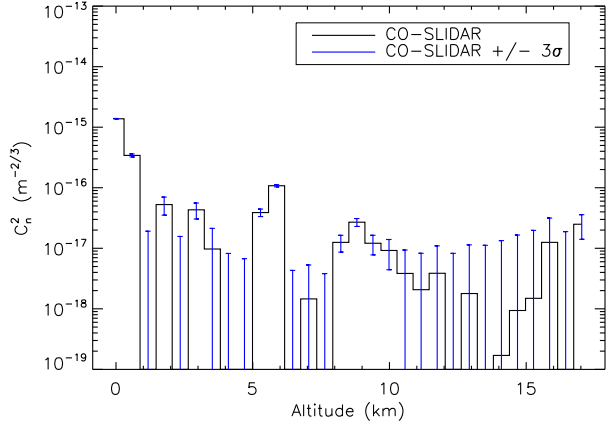}}
    \\ 
    \subfloat[]{
      \includegraphics[width=0.65\linewidth]{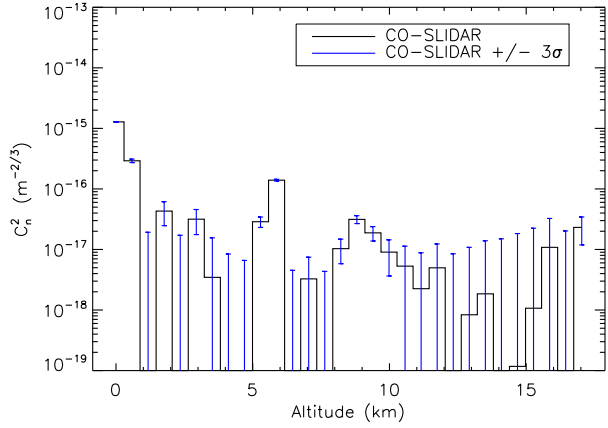}}
    \\
    \subfloat[]{
      \includegraphics[width=0.65\linewidth]{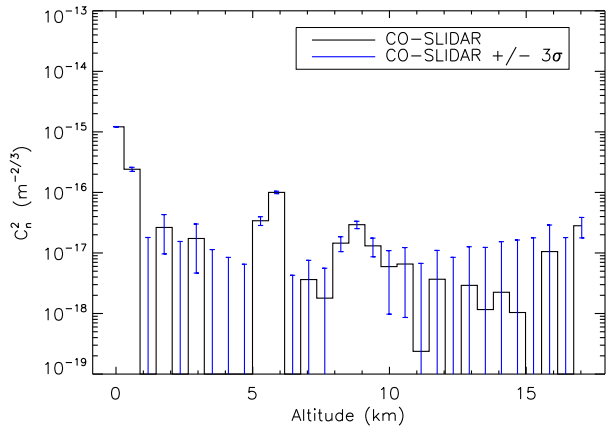}}

    \caption{(a), (b), (c) MAP CO-SLIDAR reconstructions of the $C_n^2$ profile with the estimated $3\sigma$
      error bars, for three consecutive minutes of observation. Each profile
      corresponds to one minute of observation. Data from May $15^{th}$, 2012,
      around $01 \!\! : \!\! 00$~UT.
      \label{fig:profiles_error_bars_simu}
    }  
  \end{center}
\end{figure}

Following these encouraging results we perform a MAP estimation in
order to impose some smoothness to the profile reconstruction. We use the same
regularization parameter as in the simulation case, because the $C_n^2$ profile is
of the same order of magnitude and has a similar shape. The corresponding $C_n^2$ profile
is presented in Fig.~\ref{fig:profiles_ml_map_exp} and compared to the one without regularization.
We get a slightly different profile from the ML one, smoother in the first kilometer. The differences occur mainly at low
altitudes, from $0$ to $4$~km. 
\newpage
Finally, in Fig.~\ref{fig:profiles_error_bars_simu} we add the
$3\sigma$ error bars on the reconstructed profile and we present three MAP
$C_n^2$ profiles corresponding to three consecutive minutes of observation, each profile corresponding to one minute of observation.
The three profiles are very similar. They show strong turbulence at low
altitude, another strong layer around $5$ km, and some weaker layers in
altitude. This shape of turbulence profile is typical of an astronomical site.
The mean Fried parameter estimated on the line of sight is $r_0
\simeq 6.9$~cm, and the mean variance of log-amplitude is
$\sigma_{\chi}^2 \simeq 0.045$.

\subsection{Comparison with $C_n^2$ profiles deduced from NCEP/NCAR Reanalysis}\label{sec:COSLIDAR_NCEPNCAR}
We compare the CO-SLIDAR estimation with a free atmosphere
$C_n^2$ profile deduced from NCEP/NCAR Reanalysis. This method does not allow
an estimation of the turbulence in the boundary layer, as meteorological
parameters are too unstable in this part of the atmosphere. 

\begin{figure}[htbp]
  \begin{center}\leavevmode
      \includegraphics[width=0.65\linewidth]{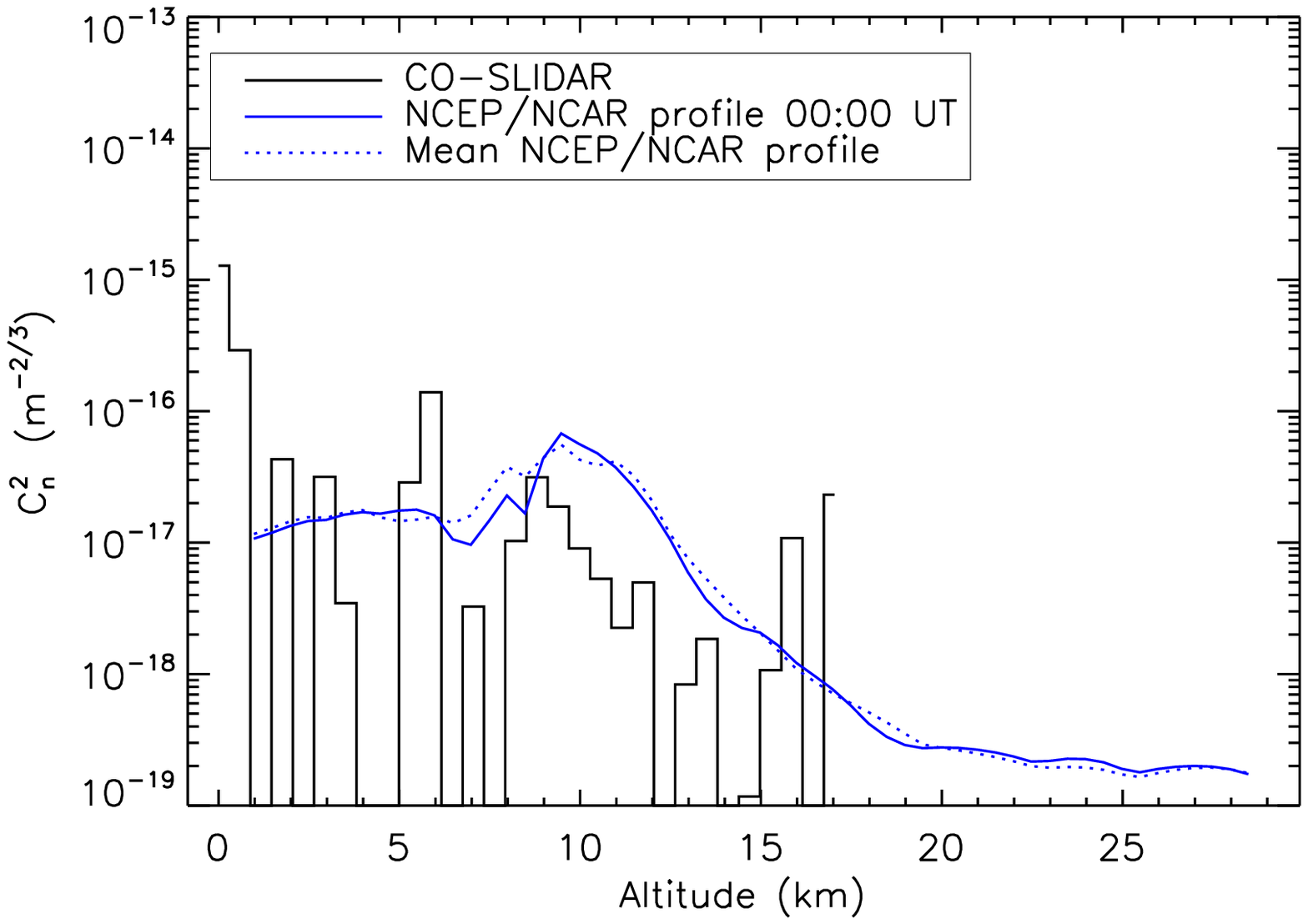}
    \caption{MAP CO-SLIDAR reconstruction of the $C_n^2$ profile. $1$ min
      dataset from May $15^{th}$, 2012, around $01 \!\! : \!\! 00$~UT.
      Comparison with the free atmosphere profile at
      $00 \!\! : \!\! 00$~UT and with the mean profile
      of May 2012, deduced from NCEP/NCAR Reanalysis.
      \label{fig:profiles_comp_COSLIDAR_NCEPNCAR}
    }  
  \end{center}
\end{figure}

As they are very similar, we only consider the second $C_n^2$ profile of those reconstructed throughout the
three consecutive minutes with the MAP estimation, around $01 \!\! : \!\! 00$~UT, on May $15^{\mathrm{th}}$,
2012. This profile is compared with the free atmosphere $C_n^2$ profile at
$00 \!\! : \!\! 00$~UT and
with the mean profile of May 2012, obtained from NCEP/NCAR Reanalysis. The
comparison is shown in Fig.~\ref{fig:profiles_comp_COSLIDAR_NCEPNCAR}.
The turbulence profiles computed from NCEP/NCAR Reanalysis are very smooth,
because they correspond to an averaging of the data, so they cannot show fast
changes in the $C_n^2$ profile, unlike the CO-SLIDAR method. 

A few differences can be noticed between the profile at $00 \!\! : \!\! 00$~h~UT and the mean profile
deduced from NCEP/NCAR Reanalysis, but their shape are very similar. 

The CO-SLIDAR $C_n^2$ profile and those obtained from NCEP/NCAR Reanalysis are
of the same order of magnitude. Between $1$ and $5$~km, the estimation from
NCEP/NCAR Reanalysis is lower than the CO-SLIDAR one, whereas between $6$ and
$15$~km it is the opposite. Both estimations show a decrease of the turbulence
strength between $9$ and $15$~km. Between $16$ and $17$~km, the CO-SLIDAR
estimation presents a rise of the turbulence strength, contrary to the
estimation from NCEP/NCAR Reanalysis. This rise with the CO-SLIDAR estimation
could correspond to the integral of unseen turbulence above $17$~km, while
this turbulence is reconstructed with the estimation from NCEP/NCAR Reanalysis. This
rise could also be an artefact because these very high altitudes
correspond to the edge of the cross-correlation maps, where the SNR is poor
and where we have few pairs of subapertures to estimate the correlations.   

The CO-SLIDAR $C_n^2$ profiles have been found to be comparable on average to
the $C_n^2$ profiles for the free atmosphere deduced from NCEP/NCAR
Reanalysis, despite the difference of the two methods, from the kind of
data they use, to the way they work.

\section{CO-SLIDAR in the $C_n^2$ profilers' landscape}\label{sec:position}
The results presented in the previous section confirm that CO-SLIDAR on meter
class telescopes provides sub-kilometric resolution $C_n^2$ profiles in the $\left[0 \;
  ; \; 20\right]$~km altitude range. This method could be used for site
characterization to obtain relevant inputs for tomographic AO design and
performance evaluation, or to help optical turbulence forecast. 

Of course, inter-comparisons are needed, with the reference profilers SLODAR,
G-SCIDAR and MASS, and with new-generation profilers, such as PML (Profiler of
Moon Limb)~\cite{ziad-a-13} and Stereo-SCIDAR~\cite{osborn-p-13}. A multi-instrument
campaign dedicated to this comparison is foreseen~\cite{masciadri-p-13}. 

Tomographic AO systems will include several wavefront sensors, leading to
multi-directional SLODARs~\cite{vidal-a-10, gilles-a-10, cortes-a-12}, but
external $C_n^2$ profilers such as CO-SLIDAR would be useful for the
calibration of these systems.

Moreover, some hardware and architecture improvements are possible on the
CO-SLIDAR instrument. The optical transmission could be increased and the sources
could be separated on two different detectors to tune the
detector gain on each star, to observe binary stars with visible magnitude up
to $6$ and thus to allow a better sky coverage. With two detectors the images' size
would be smaller and that would speed up their recording, to enable wind profiling.
Outer scale profiling is also possible. Eventually, CO-SLIDAR could be used as
an automatic monitor, by means of real time data processing.

\section{Conclusion}\label{sec:conclusion}
In this paper, we have presented the first-on sky $C_n^2$ profiles estimated
with the CO-SLIDAR profiler. The improved reconstruction method has been
described and its latest
performance has been illustrated in an end-to-end simulation, before testing
it in a real astronomical observation on the $1.5$~m MeO telescope.
$C_n^2$ profiles have been estimated with the CO-SLIDAR method and compared to
those reconstructed from correlations of slopes only or of scintillation only,
highlighting the fact that the two kinds of correlations are
complementary and allow the reconstruction of low and high altitude layers. With this CO-SLIDAR instrument, we have reconstructed $30$
turbulent layers, with a resolution of $600$~m, from the ground up to $17$~km. We have compared our results with the $C_n^2$ profiles deduced from
NCEP/NCAR Reanalysis, and we have found a good agreement between the
estimations. In a nutshell, we have shown that CO-SLIDAR is a sub-kilometric resolution $C_n^2$ profiler working on meter
class telescopes. The method now needs cross-calibrations with other optical
profilers in multi-instrument campaigns. Considered future works
include improvements to gain a better sky-coverage, performing real-time $C_n^2$
estimations, and extensions to wind and outer scale profiling.   

\section*{Acknowledgments}
This work has been performed in the framework of a Ph.D. Thesis supported by
Onera, the French Aerospace Lab, and the French Direction Générale de
l'Armement (DGA). The authors are very grateful to the team operating at MeO
station, for the use of the $1.5$~m telescope and their help and support
throughout the whole campaign. The authors also want to thank B. Fleury and F.
Mendez for help in the optical design, optical alignment and optomechanics. They
are also grateful to Y. Hach for computing the NCEP/NCAR $C_n^2$ profiles, and
to V. Michau for fruitful exchanges.

\appendix
\section{Estimation of the variance of the convergence noise for a
  Gaussian random variable}\label{sec:appendix}
In CO-SLIDAR, the correlations are estimated from a finite number of frames.
The so-called convergence noise corresponds to the difference between these empirical
estimations and the exact mathematical expectations.

In this appendix, we illustrate the methodology for computing
$C_{\mathrm{conv}}$ in the simplified case of the
estimation of the variance of a scalar Gaussian random variable by expressing
the variance of the convergence noise.

Let $X$ be a centered Gaussian random variable such as $X \sim \mathcal{N}\left(0,
  \sigma_{X}^2 \right)$ and $x_i$ a realization of $X$. An estimation of
the variance of $X$, $\sigma_{X}^2$ (which is a ML estimate because we have
assumed $X$ to be Gaussian), is given by:
\begin{equation}
V = \frac{1}{N_r}\sum_{i=1}^{N_r} x_i^2,
\end{equation}   
where $N_r$ is the number of realizations of $X$.
The convergence noise is by definition:
\begin{equation}
Y = V - \sigma_{X}^2 = \frac{1}{N_r}\sum_{i=1}^{N_r} x_i^2 - \sigma_{X}^2. 
\end{equation}
It is, in turn, a random variable. Its mean value is:
\begin{equation}
\left\langle Y \right\rangle = \frac{1}{N_r}\sum_{i=1}^{N_r} \left\langle x_i^2 \right\rangle -
\sigma_{X}^2 = 0.
\end{equation}
The variance of $Y$ can be written as:
\begin{eqnarray}\label{eq:convergence_noise_variance}
\sigma_{Y}^2 &=& \left\langle Y^2 \right\rangle - \left\langle Y \right\rangle ^2 \\
\nonumber &=& \left\langle \left(\frac{1}{N_r}\sum_{i=1}^{N_r}  x_i^2  -\sigma_{X}^2
\right)^2\right\rangle \\
\nonumber &=& \left\langle \left(\frac{1}{N_r}\sum_{i=1}^{N_r}  x_i^2
  \right)^2 \right\rangle + \sigma_{X}^4 - 2\sigma_{X}^2 \times \frac{1}{N_r}\sum_{i=1}^{N_r} \left\langle x_i^2 \right\rangle \\
\nonumber &=& \left\langle \frac{1}{N_r^2} \sum_{i,j} x_i^2x_j^2 \right\rangle  - \sigma_{X}^4 \\
\nonumber &=& \left\langle \frac{1}{N_r^2} \sum_{i=1}^{N_r} x_i^4  +
  \frac{1}{N_r^2}\sum_{i \ne j} x_i^2 x_j^2 \right\rangle - \sigma_{X}^4 \\
\nonumber  &=& \frac{1}{N_r^2}\sum_{i=1}^{N_r} \left\langle x_i^4 \right\rangle + \frac{N_r\left(N_r-1\right)}{N_r^2}\sigma_{X}^4 - \sigma_{X}^4 \\
\nonumber &=& \frac{3}{N_r}\sigma_{X}^4 -\frac{1}{N_r}\sigma_{X}^4  \\
\nonumber &=& \frac{2}{N_r} \sigma_{X}^4.
\end{eqnarray}

The variance of the convergence noise $\sigma_{Y}^2$ can therefore be deduced
from the variance of $X$. In practice the latter is unknown, so we
approximate $\sigma_{X}^2$, in the last line of
Eq.~(\ref{eq:convergence_noise_variance}), by its empirical estimate $V$.

\end{document}